\documentclass[a4paper,aps,prd,twocolumn,showpacs,eqsecnum,amsmath,amssymb,nofootinbib,superscriptaddress]{revtex4-1}
\usepackage{dcolumn}
\usepackage{bm}
\usepackage{graphics}
\usepackage{graphicx}
\usepackage{epsfig}
\usepackage{booktabs,multirow}
\usepackage{hhline}
\usepackage{slashed}
\usepackage{rccol}
\usepackage{mathrsfs}

\usepackage{xcolor}\colorlet{linkequation}{blue}
\usepackage[colorlinks=true, 
            linkcolor=red,   
            filecolor=red,   
            citecolor=green, 
            urlcolor=blue,
            hyperfootnotes=true]{hyperref}

\newcommand*{\SavedEqref}{}
\let\SavedEqref\eqref
\renewcommand*{\eqref}[1]{%
  \begingroup
    \hypersetup{
     linkcolor=linkequation,
      linkbordercolor=linkequation,
    }%
    \SavedEqref{#1}%
  \endgroup
}

\begin{document}
\newcommand{\newc}{\newcommand}
\renewcommand{\thefootnote}{\fnsymbol{footnote}}

\newc{\neutralino}{\widetilde\chi^0}
\newc{\chargino}{\widetilde\chi^{\pm}}
\newc{\squarkk}{\widetilde{q}_k}
\newc{\squarkl}{\widetilde{q}_l}
\newc{\tanb}{\tan\beta}
\newc{\gev}{\mbox{~GeV}}
\newc{\tev}{\mbox{~TeV}}

\newc{\be}{\ensuremath{\beta}}
\newc{\al}{\ensuremath{\alpha}}
\newc{\sa}{\ensuremath{\sin\alpha}}
\newc{\ca}{\ensuremath{\cos\alpha}}
\newc{\ta}{\ensuremath{\tan\alpha}}
\newc{\sbt}{\ensuremath{\sin\beta}}
\newc{\cbt}{\ensuremath{\cos\beta}}
\newc{\ma}{\ensuremath{m_{A}}}

\title{Associated production of Higgs boson with a photon at electron-positron colliders}
\author{Mehmet~Demirci}
\email{mehmetdemirci@ktu.edu.tr}
\affiliation{Department of Physics, Karadeniz Technical University, TR61080 Trabzon, Turkey}
\date{\today}
\begin{abstract}
A complete one-loop prediction for the single production of the neutral Higgs bosons in association with a photon in electron-positron collisions is presented in the framework the minimal supersymmetric standard model (MSSM), paying special attention to the individual contribution from each type of diagram. This process has no amplitude at tree level and is hence directly sensitive to one-loop impacts and the underlying dynamics of Higgs. To investigate the effect of the new physics, four different scenarios, which include a Higgs boson with mass and couplings consistent with those of the discovered Higgs boson and a considerable part of parameter space allowed by the bounds from the searches for additional Higgs bosons and sparticles, are chosen in the MSSM. The dependence of the cross section in both the standard model(SM) and MSSM on the center-of-mass energy is examined by considering the polarizations of the initial electron and positron beams. The effect of individual contributions from each type of one-loop diagram on the total cross section is also investigated in detail. Furthermore, the total cross sections of $e^- e^+ \rightarrow \gamma h^0$ as well as $e^- e^+ \rightarrow \gamma A^0$ are scanned over the plane $m_A-\tan\beta$ for each scenario. The full one-loop contributions are crucial for the analysis of beyond the SM physics at a future electron-positron collider.
\end{abstract}


\maketitle

\section{\bf Introduction}
Despite its many successes, the standard model (SM) leaves us with a lot of questions to be answered, such as the hierarchy problem, the origin of flavor, etc. Since the discovery of the Higgs boson at the LHC~\cite{Higgs_ATLAS,Higgs_CMS} and to date, no evidence of new physics beyond the SM (BSM) has yet been found. However, the observations of neutrino oscillations, matter-antimatter asymmetry, relic density of dark matter (DM), and so on open the door to new physics BSM. Additionally, there are strong motivations to extend the scalar sector of the SM by introducing more than one Higgs doublet. Therefore, the development of new attempts concentrated on the research of data which provide a hint about new physical degrees of freedom seems compulsory. This is the main goal of proposals at future $e^+ e^-$ colliders such as the International Linear Collider (ILC)~\cite{ILC1,ILC2,eeLC}, Compact Linear Collider (CLIC)~\cite{eeLC,CLIC1}, Circular Electron-Positron Collider (CEPC)~\cite{CEPC}, and Future Circular Collider (FCC)~\cite{FCC}. On the other hand, they are mainly designed to provide a high precision and complete picture of the Higgs boson and its couplings. The $e^+ e^-$ colliders compared to the hadron colliders have a cleaner background, and hence the new physics signals are easily separated from the background. The ILC is one of the most developed linear colliders planned to be a Higgs factory in the centre of mass energies of $\sqrt{s} = 250-500$ GeV (extendable up to a 1 TeV). The CLIC is a TeV-scale high luminosity linear collider planned to be operated at centre-of-mass energies of $\sqrt{s}=380$ GeV, 1.5 TeV, and 3 TeV. The CEPC collider with a circumference of 100 km is designed to operate at $\sqrt{s} = 240$ GeV. The potential for CEPC to probe a suite of loop-level corrections to Higgs and electroweak observables in supersymmetric models is comprehensively studied in Ref.~\cite{CEPC2}.

Even at $\sqrt{s} = 250\gev$ with a total integrated luminosity of $2$ ab$^{-1}$, there are proposals for the electron-positron ILC collider to precisely measure the couplings of the observed Higgs boson to gauge bosons, leptons, and quarks \cite{ILC1, ILC3} with an accuracy of order 1$\%$. 
This would allow detecting the small deviations for BSM scenarios. A very precise prediction of Higgs boson production involving additional interactions which come from BSM scenarios can provide significant hints about new physics. There are many important motivations to choose the minimal supersymmetric standard model (MSSM) as BSM scenario that could identify these new interactions.

The MSSM~\cite{Haber,Nilles,Gunion,wss}, one of the most attractive and widely considered extensions of SM, keeps the  number of new fields and couplings to a minimum. It provides a solution for the hierarchy problem of the SM and offers a candidate for the DM postulated to explain astrophysical observations and a prediction for the mass of the scalar resonance observed at the LHC. The MSSM has two Higgs doublets, which lead to a physical spectrum that includes a couple of charged Higgs bosons $H^\pm$, a \textit{CP}-odd Higgs boson $A^0$, and the light/heavy \textit{CP}-even Higgs bosons $h^0$/$H^0$ in the \textit{CP}-conserving case. In the Higgs sector of the MSSM, all couplings and masses at tree level can be described by only two parameters: the mass of pseudoscalar Higgs boson $m_{A^0}$ and the ratio of the vacuum expectation values of the two doublets $\tan\beta$. The discovered Higgs with mass of around 125 GeV could be interpreted naturally as one of the two neutral \textit{CP}-even Higgs bosons $h^0$ or $H^0$ in the MSSM\footnote{In particular, the mass of the light Higgs boson $h^0$ is bounded from above by $m_Z |\cos2\beta|$ at tree level. However, radiative corrections can significantly change the tree level Higgs mass predictions, allowing for $m_h \approx 125 \gev$, which is compatible with the observed Higgs boson.}~\cite{Heinemeyer,Scopel,Djouadi}. Moreover, many new particles in the MSSM are predicted such as scalar leptons $\widetilde{l}$, scalar quarks $\widetilde{q}_k$, neutralinos $\neutralino_{i}$, and charginos $\chargino_j$. In R parity-conserving models~\cite{Fayet}, supersymmetric particles (or sparticles, for short) are pair produced, and their decay chains end in the stable, lightest sparticle (LSP). The lightest neutralino $\neutralino_{1}$ is considered as a LSP in many models. As a neutral, weakly interacting, and stable particle, $\neutralino_{1}$ is consistent with the properties required of a DM candidate~\cite{PDG}.

The associated Higgs production with a photon, $e^- e^+ \rightarrow \gamma h^0$, is well suited to study the Higgs to neutral gauge boson couplings such as the $h\gamma\gamma$ and $hZ\gamma$ couplings. Future $e^+ e^-$ colliders are optimal for studying $e^- e^+ \rightarrow h \gamma$, where the cross section in the SM~\cite{Barroso, Abbasabadi, Djouadi2} has a peak about $\sqrt{s} = 250\gev$. Because the tree-level contribution of the process is highly suppressed by the electron mass and the process is protected by the electromagnetic gauge symmetry, it occurs at the one-loop level for the first time. Therefore, the size of cross section is an order of magnitude of $10^{-1}$ fb at $\sqrt{s} =$ 250 GeV, which is rather small. However, since the signal is very clean, i.e., a monochromatic photon in the final state, it may be observed at the above future $e^+ e^-$ colliders with the design luminosity. Furthermore, new physics contributions can considerably increase the rate of production relative to the SM case; namely, the process is potentially very sensitive to new physics.

There are several studies dedicated to the investigations of new physics effects on the process in the framework of an effective field theory or anomalous Higgs-boson couplings~\cite{anomalous,anomalous2,eft} as well as the extended Higgs models [inert doublet/triplet model and two Higgs doublet model (THDM)]~\cite{IDM,extendedHiggsmodel} and the MSSM \cite{Djouadi2,MSSM,MSSM2,CMSSM}. The enhancement effects in the process $e^+e^- \to h\gamma$ and its correlation with the LHC signal strengths $R_{\gamma\gamma}$ and $R_{Z\gamma}$ in the MSSM have been examined in Ref.~\cite{MSSM}. The ``supersimple" expressions for all helicity amplitudes of the same process have been derived in Ref.~\cite{MSSM2}. In Ref.~\cite{CMSSM}, the authors investigate the neutral Higgs boson production in association with another Higgs boson or a SM gauge boson at $e^+e^-$ colliders in the framework MSSM with complex parameters, taking into account soft and hard QED radiation. In Ref.~\cite{Agamma}, a \textit{CP}-odd Higgs boson $A^0$ in association with a photon are studied in the MSSM with \textit{CP}-violating phases. However, owing to the most recent constraints on the parameter space of the MSSM from Run 2 of the LHC,  the size of the MSSM contributions to $e^- e^+ \rightarrow h \gamma$ should be reevaluated in the allowed parameter space.

In this work, the single production of the neutral Higgs bosons in association with a photon in electron-positron collisions is reinvestigated in the framework MSSM, paying special attention to the individual contribution from each type of diagram. In this aim, it is examined that how and how much the individual contributions from each type of diagram could amplify or lessen the $h^0 \gamma$ signal at a future $e^- e^+$ collider. For this aim, four different benchmark scenarios, which have a Higgs boson with mass and couplings consistent with those of the discovered Higgs boson and a considerable part of their parameter space is allowed by the bounds from the searches for additional Higgs bosons and supersymmetric particles, are chosen. These scenarios are named $m_h^{125}$, $m_h^{125}(light~\widetilde\tau)$, $m_h^{125}(light~\widetilde\chi)$, and $m_h^{125}(alignment)$~\cite{Carena, mh125}. Distributions for the total cross sections are computed as a function of the center-of-mass energy and the polarization of the incoming beams. Furthermore, the total cross sections of both $e^- e^+ \rightarrow \gamma h^0$ and $e^- e^+ \rightarrow \gamma A^0$ are scanned over the plane $m_A-\tan\beta$ for each scenario, and regions in which the cross section is large enough to be detectable at a future $e^- e^+$ collider are determined in this paper.

The contents of the present work are the following. Section~\ref{sec:cros} provides the Feynman diagrams and the analytical expressions related to the process $e^- e^+ \rightarrow h \gamma$. This section then gives information on how the numerical evaluation is done. Section \ref{sec:input} provides details of the considered benchmark scenarios. In Sec. \ref{sec:results}, numerical results are presented, and the some parameter dependencies of the cross section are discussed in detail. Finally, Section~\ref{sec:conclusion} presents the conclusions of the present study.

\section{Theoretical Framework}\label{sec:cros}
\begin{figure*}[!tbph]
    \begin{center}
\includegraphics[width=1\linewidth]{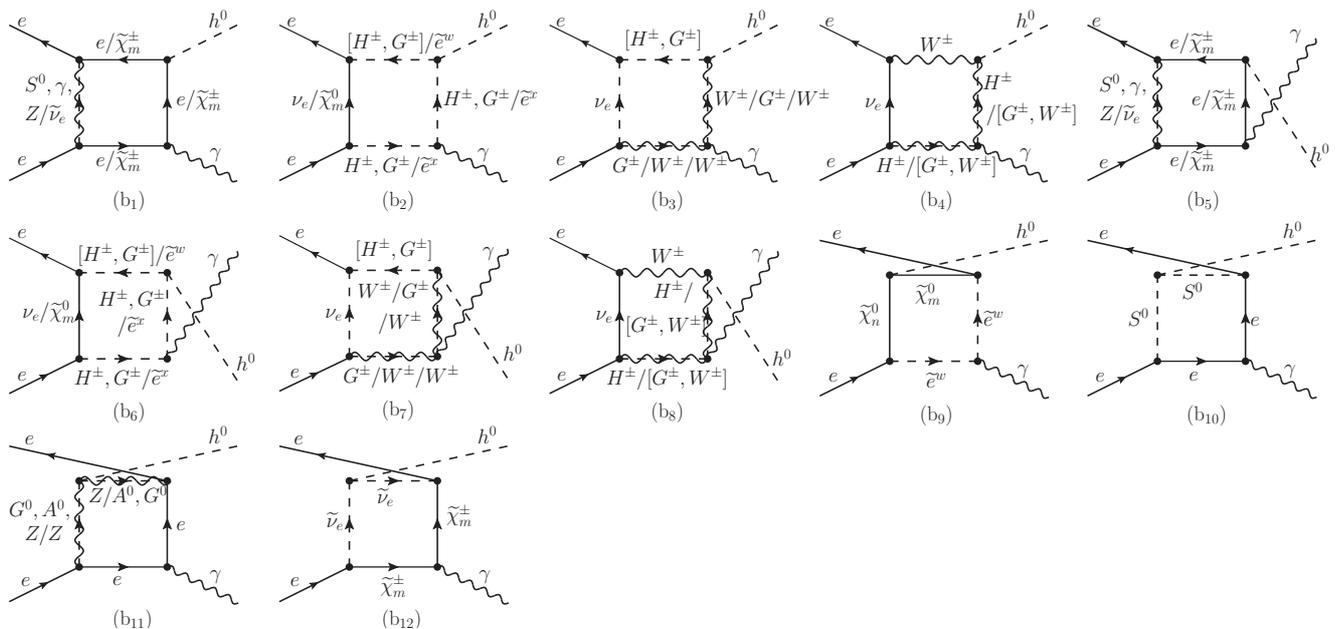}
     \end{center}
     \vspace{ -4mm}
\caption{Box-type diagrams contributing to the process $e^+ e^- \rightarrow h \gamma$
at one-loop level. }\label{fig:box}
\end{figure*}
\begin{figure*}[!ht]
    \begin{center}
\includegraphics[width=1\linewidth]{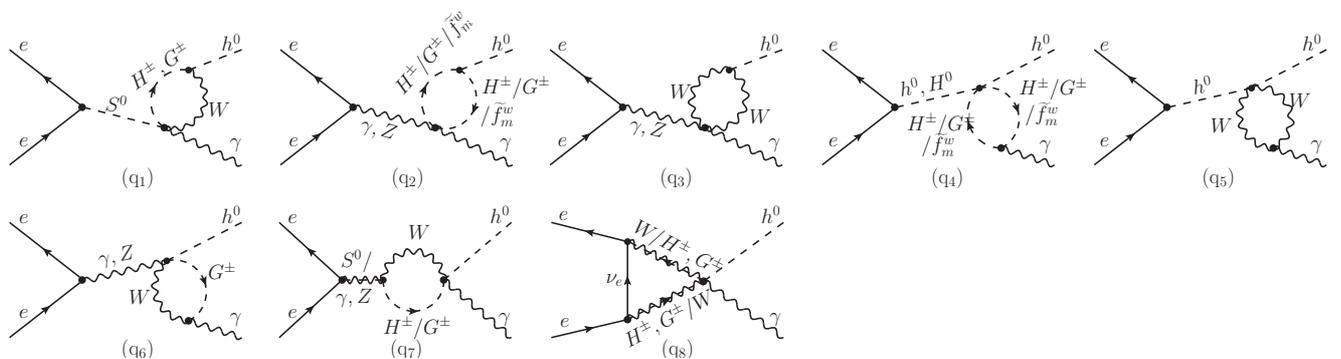}
     \end{center}
     \vspace{ -2mm}
\caption{Quartic interaction diagrams contributing to the process $e^+ e^- \rightarrow h \gamma$
at one-loop level.}\label{fig:qua}
\end{figure*}
\begin{figure*}[!ht]
    \begin{center}
\includegraphics[width=1\linewidth]{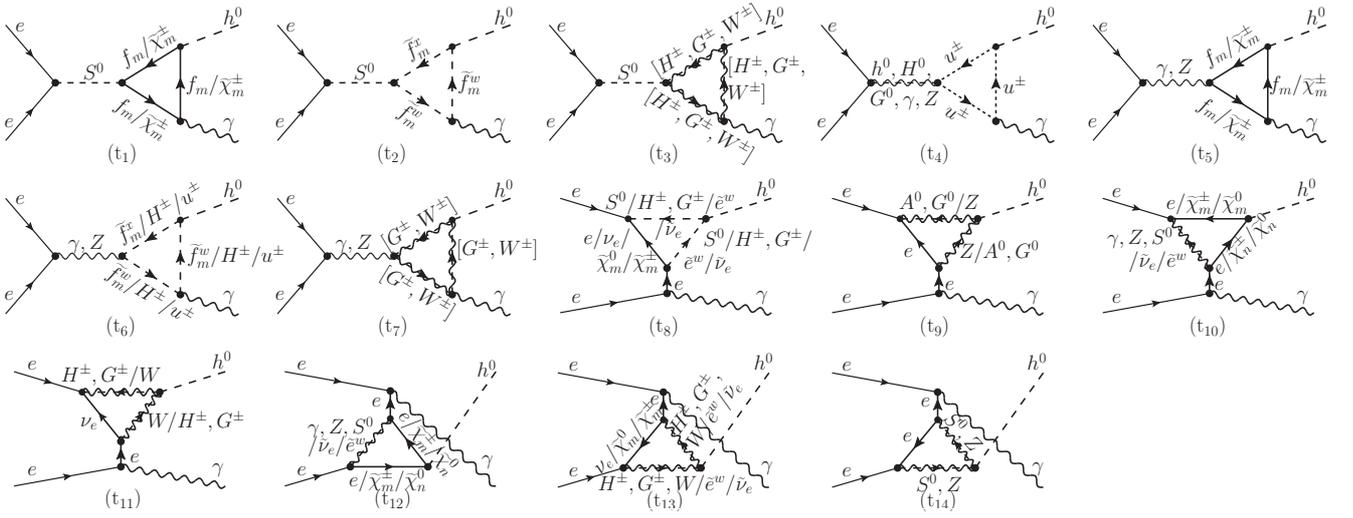}
     \end{center}
     \vspace{ -2mm}
\caption{Triangle-type diagrams contributing to the process $e^+ e^- \rightarrow h \gamma$ at one-loop level.} \label{fig:tria}
\end{figure*}
\begin{figure*}[!ht]
    \begin{center}
\includegraphics[width=1\linewidth]{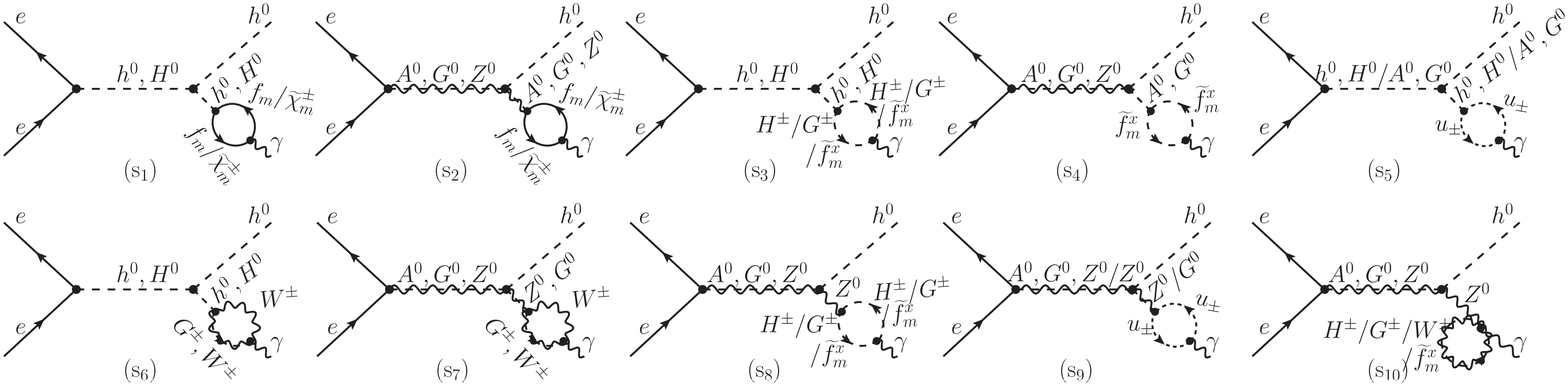}
     \end{center}
     \vspace{ -2mm}
\caption{Self-energy diagrams contributing to the process $e^+ e^- \rightarrow h \gamma$ at one-loop level.} \label{fig:self}
\end{figure*}

The associated production of single Higgs boson with a photon in an electron-positron collision is indicated by
\begin{equation} \label{eq:gammah}
e^+(p_1) e^-(p_2) \rightarrow h(k_1) \gamma(k_2),
\end{equation}
where after each particle, as usual, its 4-momenta is written in parentheses. The Mandelstam variables can be written as
\begin{equation}
s=(p_1+p_2)^2, \quad t=(p_1-k_1)^2,\quad
u=(p_1-k_2)^2.
\end{equation}
At tree level, the process occurs via the $t$-channel electron-exchange diagram suppressed by the mass of the electron. Its amplitude is neglected at tree level, i.e., for the first time, it is mediated by one-loop diagrams, and hence it is sensitive to all virtual particles inside the loop.

The total amplitude of one-loop level can be written as a linear sum of box, triangle, and bubble one-loop integrals. According to the type of loop correction, the one-loop diagrams contributing to the process $e^+ e^- \rightarrow h \gamma$ can be classified into four different types: the box-type, self-energy, quartic coupling-type, and triangle-type diagrams. A complete set of one-loop Feynman diagrams and the corresponding amplitudes for both SM and MSSM are generated by the \textsc{FeynArts} \cite{Feynarts}. For MSSM, the diagrams\footnote{The diagrams were drawn by using~\textsc{JaxoDraw}~\cite{JaxoDraw,JaxoDraw2}.} are shown in Figs.~\ref{fig:box} to~\ref{fig:tria}. There is also another set of diagrams in which particles in each loop are running counterclockwise. The brackets $[\ldots]$ represent that all possible combinations of the particles in the brackets can be written. In the Feynman diagrams, the label ${f}_m~(\widetilde{f}^x_m)$ refers to fermions (scalar fermions) $e_m,u_m,d_m~(\widetilde{e}^x_m, \widetilde{u}^x_m, \widetilde{d}^x_m)$, and the label $S^0$ represents all neutral Higgs/Goldstone bosons $h^0, H^0, A^0, G^0$. The indices $m$ and $x$ represent the generation of (scalar-)quark and the scalar-quark mass eigenstates, respectively. 
In loop diagrams, scalar particles such as neutral and charged Higgs/Goldstone bosons, sfermions, are denoted by dashed lines, and $\gamma$, $Z$, and $W$ bosons are denoted by wavy lines.

Figure~\ref{fig:box} shows all possible box-type contributions, which have the loops of neutrino, electron, selectron $\widetilde{e}^{1,2}$, charginos $\widetilde\chi^\pm_{1,2}$, neutralinos $\widetilde\chi^0_{1,2,3,4}$, neutral Higgs bosons ($h^0$, $H^0$, $A^0$, $G^0$), $Z$ boson, $W$-boson and charged Higgs/Goldstone bosons ($H^\pm$, $G^\pm$).

Figure~\ref{fig:qua} shows all quartic coupling-type diagrams which consist of bubbles (q$_{1-7}$) attached to the initial state via an intermediate $\gamma$ or $Z$ or neutral Higgs bosons ($h^0$, $H^0$, $A^0$, $G^0$) and the triangle loop (q$_{8}$) of the neutrino $\nu_e$, charged Higgs/Goldstone boson, and $W$-boson directly attached to the final state.

Figure~\ref{fig:tria} presents all triangle-type contributions which consist of triangle vertices (t$_{1-7}$) attached to the initial state via an intermediate $\gamma$ or $Z$ or neutral Higgs bosons and also includes triangle corrections to $t$-channel electron exchange.

Finally, Fig.~\ref{fig:self} shows self-energy diagrams contributing to $e^+ e^- \rightarrow h \gamma$. The self-energy contributions appear from diagrams in which the Higgs boson $h^0$ is emitted from the virtual $S^0$ or $Z$ line and with $Z-\gamma$ mixing via a fermion or $W$ boson or charged Higgs boson or chargino or sfermion loops and diagrams with the mixing between the neutral Higgs bosons and the photon $\gamma$. Note that diagrams with $S^0-\gamma$ mixing and with the virtual $S^0$ line give negligible contributions. There are also $t$-channel self-energy diagrams with the electron exchange, but they are highly suppressed by the electron mass; they are not explicitly shown here. Consequently, the main contribution comes from $Z-\gamma$ mixing diagrams with the virtual $Z$ line.

The $Z-\gamma$ mixing diagrams are needed in order to obtain the finite results. The fermionic contributions to triangle-type ($t_{1-7}$) and $Z-\gamma$ mixing self-energy diagrams are gauge invariant by themselves, but the $W$ contributions are not. Gauge invariance of the $W$ contributions is supplied only when the box-type and $t$-channel triangle diagrams are included~\cite{Djouadi2}. In all Feynman diagrams computed here, there is no virtual photon in the loops; hence, the results are infrared finite. Real or virtual emission of the photon is suppressed by the electron mass.
\begin{table*}[!bht]
\caption{Triple and quartic Higgs couplings and couplings of the Higgs bosons to gauge bosons, which are included in each type of diagram. Here, S, V and F refer to scalar, vector, and fermion, respectively.}\label{tab:couplings}
\begin{tabular}{|c|l|c|c|c|c|c|}
\hline
&\textbf{~~~~~~Couplings}&\textbf{Box type} & \textbf{Triangle type} & \textbf{Bubble type}  &\textbf{Quartic type}  &\textbf{Self energy} \\
\hline\hline
\parbox[t]{2mm}{\multirow{8}{*}{\rotatebox[origin=c]{90}{SSS}}} &$\lambda_{h^0 [h^0,H^0] [h^0,H^0] }$   &\checkmark (b$_{10}$) &\checkmark (t$_{8,14}$)&     &      &\checkmark (s$_{1,3,5,6}$) \\
&$\lambda_{h^0 [A^0,G^0] [A^0,G^0] }$   &\checkmark (b$_{10}$) &\checkmark (t$_{8,14}$)&     &      &\checkmark (s$_{2,4,5,7}$) \\
&$\lambda_{h^0 H^+ H^- }$   &\checkmark (b$_{2,6}$)            &\checkmark (t$_{3,6,13}$)&\checkmark (q$_{2}$)&    &\checkmark (s$_{3}$)    \\
&$\lambda_{h^0 G^+ H^- }$   &\checkmark (b$_{3,6,7}$)            &\checkmark (t$_{3,8,13}$)&          &    &   \\
&$\lambda_{h^0 G^+ G^- }$   &\checkmark (b$_{3,6,7}$)            &\checkmark (t$_{3,6,7,8,13}$)&\checkmark (q$_{2}$)&    &\checkmark (s$_{3}$)   \\
&$\lambda_{H^0 H^-G^+}$   &           &\checkmark (t$_3$) &        &      & \\
&$\lambda_{H^0 G^-G^+}$   &           &\checkmark (t$_3$) &        &     &\checkmark (s$_{3}$)  \\
&$\lambda_{A^0 H^-G^+}$   &           &\checkmark (t$_3$) &        &      & \\
\hline
\multirow{6}{*}{\rotatebox[origin=c]{90}{\parbox[c]{1cm}{\centering VSS}}} &$\lambda_{h^0 H^+W^- }$          &\checkmark (b$_{3,4,7,8}$)& \checkmark (t$_{3,11}$)          &\checkmark (q$_{1}$)     &      & \\
&$\lambda_{h^0 G^+W^- }$          &\checkmark (b$_{3,4,7,8}$)& \checkmark (t$_{3,7,11}$)          &\checkmark (q$_{1}$)     &   & \\
&$\lambda_{H^0 [G^+,H^+]W^- }$          &              & \checkmark (t$_{3}$)          &     &   & \\
&$\lambda_{A^0 H^+ W^-}$          &              & \checkmark (t$_{3}$)          &     &    &\\
&$\lambda_{G^0 G^+ W^-}$          &              & \checkmark (t$_{3}$)          &     &    &\\
&$\lambda_{h^0 [A^0,G^0] Z }$     & \checkmark (b$_{11}$)          &\checkmark (t$_{9}$)             &       &    &\checkmark (s$_{2,4,7,8,9,10}$)   \\
&$\lambda_{[\gamma,Z] H^+ H^-}$   &              &\checkmark (t$_{6}$)&\checkmark (q$_{4}$)   &    &\checkmark (s$_{8}$)   \\
&$\lambda_{[\gamma,Z] G^+ G^-}$   &              &\checkmark (t$_{7}$)&\checkmark (q$_{4}$)   &      &\checkmark (s$_{8}$)  \\
\hline
\multirow{3}{*}{\rotatebox[origin=c]{90}{\parbox[c]{1cm}{\centering VVS}}} &$\lambda_{h^0 Z Z }$     & \checkmark (b$_{11}$)          &\checkmark (t$_{9}$)    &       &       &\checkmark (s$_{2,7,8,9,10}$) \\
&$\lambda_{h^0 W W}$   &\checkmark (b$_{3,4,8}$)            &\checkmark (t$_{3,7,11,13}$)&\checkmark (q$_{3}$)&     &\checkmark (s$_{6}$)   \\
&$\lambda_{H^0 W W}$   &              &\checkmark (t$_{3}$)&          &      &\checkmark (s$_{6}$) \\
&$\lambda_{[\gamma,Z] G^+ W}$   &              &\checkmark (t$_{7}$)&\checkmark (q$_{6}$)   &     &\checkmark (s$_{7}$)  \\
\hline
\multirow{10}{*}{\rotatebox[origin=c]{90}{\parbox[c]{1cm}{\centering SSSS/VVSS}}}&$\lambda_{[h^0,H^0] h^0 H^+H^- }$     &                      &             &\checkmark (q$_{4}$) &       &  \\
&$\lambda_{[h^0,H^0] h^0 G^+G^- }$     &                      &             &\checkmark (q$_{4}$) &      &   \\
&$\lambda_{h^0 h^0 W W}$     &                      &             &\checkmark (q$_{5}$) &     &    \\
&$\lambda_{ [\gamma,Z] H^+ H^- \gamma}$     &                      &             &\checkmark (q$_{2}$) &      &\checkmark (s$_{10}$)   \\
&$\lambda_{ [\gamma,Z] G^+ G^- \gamma}$     &                      &             &\checkmark (q$_{2}$) &      &\checkmark (s$_{10}$)   \\
&$\lambda_{h^0 [H^+,G^+] W\gamma}$     &                      &             &\checkmark (q$_{1,6,7}$) &\checkmark (q$_{8}$)& \\
&$\lambda_{[H^0,A^0,G^0] [H^+,G^+] W\gamma}$     &             &             &\checkmark (q$_{1}$) &      &   \\
&$\lambda_{Z h^0 W G^+}$     &                      &             &\checkmark (q$_{6}$) &  &\\
\hline
\end{tabular}
\end{table*}

In this study, the process $e^+ e^- \rightarrow A^0 \gamma$ is also examined. Because of the \textit{CP} nature of the pseudoscalar Higgs boson $A^0$, the process $e^+ e^- \rightarrow A^0 \gamma$ has no $W$ and $Z$ contributions in the box-type diagrams, and no contribution from $Z$ boson, $W$ boson, $\widetilde{f}^x_m$, and $H^\pm$ in the triangle-type and bubble-type diagrams, compared to the process  $e^+ e^- \rightarrow h^0 \gamma$. 
There are no $W$-loop contributions because the pseudoscalar Higgs boson $A^0$ does not couple to the $W$ boson. The process $e^+ e^- \rightarrow A^0 \gamma$ receives contribution only from fermion loops (SM fermions and also supersymmetry (SUSY) fermions). Note that sfermions do not contribute to $e^+ e^- \rightarrow A^0 \gamma$ due to the fact that $A^0 \widetilde{f}_1 \widetilde{f}_2^* = A^0 \widetilde{f}_2 \widetilde{f}_1^*$.

The Higgs sector of the MSSM at tree level is described by two parameters, $\tan\beta$ and a mixing angle $\alpha$ in the \textit{CP}-even Higgs sector. Furthermore, the angle $\alpha$ could be also given in terms of $m_A$ and $\tan\beta$ as follows:
\begin{equation} \label{eq:tan2al}
\tan(2\alpha)=\tan(2\beta) \frac{m_A^2+m_Z^2}{m_A^2-m_Z^2}, ~~~-\frac{\pi}{2}<\alpha<0 .
\end{equation}
Once $m_A$ and $\tan\beta$ are given and the leading radiative correction is involved in $\alpha$, all the couplings of the Higgs bosons are fixed. Table~\ref{tab:couplings} lists the couplings of the neutral Higgs boson to gauge bosons and Higgs bosons which are included in each type of diagram for the process $e^+ e^- \rightarrow h^0 \gamma$. The Feynman diagrams are dominated by triple couplings $\lambda_{h^0 G^+ H^-}$,  $\lambda_{h^0 H^+W^-}$, $\lambda_{h^0 G^+W^-}$, and $\lambda_{h^0 G^+ G^-}$, which are proportional to mixing angles $\cos({\beta-\alpha})$ and  $\sin({\beta-\alpha})$. Some couplings of the Higgs boson are given by\footnote{The short-hand notations $s_x$ and $c_x$ are used for
$\sin(x)$ and $\cos(x)$, respectively. For example, $s_{\al+\be}=sin(\al+\be)$ for $x=\al+\be$.}
\begin{equation}\label{eq:lambda1}
\lambda_{h^0 h^0 h^0}^{MSSM} = -\frac{3i g  m_W }{2 c_W^2} c_{2\al} s_{\be + \al},
\end{equation}
\begin{equation}\label{eq:lambda2}
\lambda_{h^0 h^0 H^0}^{MSSM} = \frac{i g m_W}{2c_W^2}\bigg[ c_{2\al} c_{ \al +\be} -2 s_{2\al} s_{\al+ \be}\bigg],
\end{equation}
\begin{equation}\label{eq:lambda3}
\lambda_{h^0 H^- H^+}^{MSSM} = -\frac{i g m_W}{2}\bigg[ \frac{c_{2\be} s_{\be + \al}}{c_W^2}+2s_{\be - \al}\bigg],
\end{equation}
\begin{equation}\label{eq:lambda4}
\lambda_{h^0 G^+ H^-}^{MSSM} = -\frac{i g m_W}{2}\bigg[ \frac{s_{2\be} s_{\al + \be}}{c_W^2}-c_{\be - \al}\bigg],
\end{equation}
\begin{equation}\label{eq:lambda5}
\lambda_{h^0 G^- G^+}^{MSSM} = \frac{i g  m_W}{2 c_W^2}c_{2\be}s_{\al + \be},
\end{equation}
\begin{equation}\label{eq:lambda6}
\lambda_{h^0 A^0 A^0}^{MSSM} =-\frac{i g m_W}{2 c_W^2}c_{2\be} s_{\al + \be},
\end{equation}
\begin{equation}\label{eq:lambda7}
\lambda_{h^0 H^- W^+}^{MSSM} = -\frac{i g}{2} c_{\be-\al},
\end{equation}
\begin{equation}\label{eq:lambda8}
\lambda_{h^0 G^- W^+}^{MSSM} = -\frac{i g }{2} s_{\be-\al},
\end{equation}
\begin{equation}\label{eq:lambda9}
\lambda_{h^0 W^- W^+}^{MSSM} = i g m_W s_{\be-\al},
\end{equation}
\begin{equation}\label{eq:lambda10}
\lambda_{h^0 h^0 H^- H^+}^{MSSM} = -\frac{i g^2 }{4} \bigg[ 1+ \frac{c_{2\al} c_{2\be}s_W^2}{c_W^2}-s_{2\al} s_{2\be}\bigg],
\end{equation}
\begin{equation}\label{eq:lambda11}
\lambda_{h^0 H^- \gamma W^+}^{MSSM} = \frac{i g^2 s_W}{2} c_{\be-\al}.
\end{equation}
where the gauge coupling constant $g=e/s_W$ and $m_W$ is the mass of $W$ boson. All these couplings have a strong dependence on the mixing angles $\alpha$ and $\beta$. This study, in particular, is interested in triple Higgs couplings and couplings of the scalar to gauge bosons.

The tree-level amplitudes of process $e^+ e^- \rightarrow h^0 \gamma$ are suppressed by the electron
mass. Therefore, the process has only one-loop contributions as the lowest order, and its one-loop amplitude can be easily obtained by summing all unrenormalized reducible and irreducible contributions. Consequently, the finite and gauge-invariant results are obtained. Note that the contributions from the $t$-channel diagrams $t_{8-14}$ with electron exchange include terms proportional to $ln(m_e^2)$; thus, infrared singularity appears. To deal with this singularity, the electron mass is set a finite value. After adding the $W$-$/Z$-box contributions, this dependence on $ln(m_e)$ disappears.

The corresponding total amplitude can be given in the form
\begin{equation}\label{eq:totalM}
{\cal M}= {\cal M}_{\triangle} + {\cal M}_{\Box}+ {\cal M}_{\bigcirc}
\end{equation}
as a sum over all contributions from triangle, box, and self-energy diagrams. The differential cross section of the process, summing over the polarization of the photon, can be calculated by
\begin{equation} \label{eq:difsigma}
\frac{d\sigma}{d\cos\theta}(e^+ e^- \rightarrow h^0 \gamma)=\frac
{s-m_h^2}{32\pi s^{2}} \sum_{pol} |{\cal M}|^2,
\end{equation}
where $\sqrt{s}$ are the center-of-mass energy of $e^+e^-$ collisions and $\theta$ is the scattering angle between the photon and the electron in the centre-of-mass frame. The integrated cross section over all $\theta$ angles is given by
\begin{equation}  \label{eq:total_cross}
\sigma(e^+ e^- \rightarrow h^0 \gamma)=\int_{-1}^{+1} d\cos\theta \frac{d\sigma}{d\cos\theta}.
\end{equation}
At an $e^+ e^-$ collider, the photon in association with the Higgs is produced as monochromatic with an energy of
\begin{equation}
E_\gamma=\frac{s-m_h^2}{2\sqrt{s}}.
\end{equation}
At $\sqrt{s} = 250$ GeV, this gives a ``spectral line" at $E_\gamma=93.75$ GeV. The signal is easy to separate from the backgrounds.

With the help of \textsc{FeynArts}\cite{Feynarts} and \textsc{FormCalc}\cite{Hahn} packages\footnote{Using the same tools, we have previously done a few more recent studies~\cite{Demirci14, Demirci16, Demirci19} and achieved significant results.}, the computation is carried out in the 't Hooft-Feynman gauge.
The corresponding amplitudes are generated by \textsc{FeynArts}. The analytical results of the squared amplitude are provided by \textsc{FormCalc}. The scalar integrals in loop amplitudes are evaluated by \textsc{LoopTools}~\cite{loop}. The integration over phase space of $2 \rightarrow 2$ is numerically evaluated by using the CUBA library. This library provides routines for multidimensional numerical integration. The properties of MSSM Higgs bosons are obtained by using \textsc{FeynHiggs} \cite{FeynHiggs}.

To regulate the ultraviolet divergence (UV) in the virtual corrections, we adopt the constrained differential renormalization~\cite{CDR}, which has been shown to be equivalent to dimensional reduction at one-loop level~\cite{DR,DR2}. Hence, a supersymmetry-preserving regularization scheme is supplied by the implementation given in Ref.~\cite{DR3}. For the gamma-$Z$ mixing contribution, the on-shell renormalization scheme is used. In the on-shell renormalization scheme such as Ref.~\cite{Arhrib04}, there are no counter-terms for $Z\gamma h$ and $\gamma\gamma h$. As a result, the total amplitude is UV finite, and this can be checked numerically by varying the parameters that regularize the infinity in the loop integrals. Also, it is clear that our results are consistent with those of the SM obtained in the previous studies (see, e.g. Refs.~\cite{IDM,extendedHiggsmodel}).

The polarization effects are significant at electron-positron colliders and can be used to confer important advantages. In this study, the effect of beam polarizations on the cross section is also analyzed. Since the
electron and positron have only two spin states, the cross section for general beam polarizations is defined by~\cite{ILC1}
\begin{equation}  \label{eq:polsigma}
\begin{split}
\sigma_{P_{ e^-}P_{ e^+}}=&\frac{1}{4}\big[( 1-P_{ e^-})(1-P_{ e^+})\sigma_{LL}\\
&+  ( 1+P_{ e^-})(1+P_{ e^+})\sigma_{RR} \\
&+( 1-P_{ e^-})(1+P_{ e^+})\sigma_{LR}\\
&+( 1+P_{ e^-})(1-P_{ e^+})\sigma_{RL}\big],
\end{split}
\end{equation}
where $P_{ e^-}$/$P_{ e^+}$ indicates to the longitudinal polarizations of the initial electron/positron beam, equal to $-1$ (+1) for the completely polarized left-(right)-handed beam. $\sigma_{LL}$, $\sigma_{RR}$, $\sigma_{LR}$, and $\sigma_{RL}$ indicate the cross sections with completely polarized beams of the four possible cases. At $s$-channel $e^-e^+$ annihilation processes, only $\sigma_{LR}$ and $\sigma_{RL}$ are nonzero under the condition of helicity conservation. The intrinsic left-right asymmetry of the cross section can be calculated with
\begin{equation}  \label{eq:polsigma}
A_{LR}=\frac{\sigma_{LR}-\sigma_{RL}}{\sigma_{LR}+\sigma_{RL}}
\end{equation}
for a given process.

\begin{table}[b]
\caption{The centre of mass energies $\sqrt{s}$, integrated luminosities $\mathscr{L}$, and beam polarizations $P$, proposed at the future $e^+e^-$ colliders ILC~\cite{ILC1}, CLIC~\cite{CLIC1}, CEPC~\cite{CEPC}, and FCC$_{\rm ee}$~\cite{FCC}.}
\label{tab:collider}
\begin{center}
\begin{tabular}{|l|c|c|c|}
\hline
\rule{0pt}{1.2em}%
Collider &  $\sqrt{s}$ (GeV) & $\mathscr{L}$ (ab$^{-1}$) & ($P_{ e^-}$,$P_{ e^+}$) ($\%$) \\ [2pt]\hline\hline
ILC$_{250}$ & $250$ & $2.0$& \multirow{2}{*}{($\pm80\%,\mp30\%$)} \\
ILC$_{500}$ & $250+500$ & $6.0$& \\
\hline
CLIC$_{380}$ & $380$ & $1.0$ & \multirow{3}{*}{($\pm80\%,0$)}\\
CLIC$_{1500}$& $380+1500$ & $2.5$ & \\
CLIC$_{3000}$ & $380+1500+3000$ & $5.0$ & \\
\hline
CEPC & $240$ & $5.6$ &($0,0$)\\
\hline
FCC$_{{\rm ee}_{240}}$ & $240$ & $5.0$ & \multirow{2}{*}{($0,0$)} \\
FCC$_{{\rm ee}_{365}}$ & $240$+365 & $6.5$ &  \\
\hline
\end{tabular}
\end{center}
\end{table}
For the purpose of discussing the expected accuracy of future measurements, the centre of mass energies, integrated luminosities $\mathscr{L}$, and beam polarizations, proposed at the future electron-positron colliders are presented in Table~\ref{tab:collider}.

\section{Definition of the Benchmark Scenarios in MSSM}\label{sec:input}
This section provides details of the benchmark scenarios considered in the present study. The four different benchmark scenarios, which are called $m_h^{125}$, $m_h^{125}(light~\widetilde\tau)$, $m_h^{125}(light~\widetilde\chi)$, and $m_h^{125}(alignment)$, proposed in Refs.~\cite{Carena, mh125}, are used to illustrate the effect of the new physics based on the MSSM. Note that all of the considered scenarios include a \textit{CP}-even Higgs boson with mass about 125 GeV and couplings consistent with those of the discovered Higgs boson, and a considerable part of their parameter space is allowed by the bounds from the searches for additional Higgs bosons and supersymmetric particles.
In particular, they are consistent with the most recent experimental results from the LHC-Run 2.
The lighter \textit{CP}-even Higgs $h^0$ of the MSSM is SM-like in all scenarios. In each scenario, two free parameters are left: $\tan\beta$ and $m_A$. Hence, for each scenario, the cross section can be presented in the plane of $m_A-\tan\beta$. The parameters $\tan\beta$ and $m_A$ are varying in the range of $100\gev \leq m_A \leq2\tev$ and $0.5 \leq \tan\beta \leq 50$ for the first three scenarios and $200\gev \leq m_A \leq2\tev$ and $1 \leq \tan\beta \leq 20$ for the last scenario.

Considering the bounds placed on SUSY parameters from current experimental results, especially the direct SUSY research results from ATLAS \cite{sqgl_ATLAS17,sqgl_ATLAS18} and CMS experiments \cite{sqgl_CMS17,sqgl_CMS172,sqgl_CMS18}, a common soft SUSY-breaking mass parameter is fixed as $M_{\widetilde{f}}=2\tev$ for the first and second generations in the slepton and squark sector. The remaining parameters, which are the  gaugino mass parameters $M_1$, $M_2$, and $M_3$; the Higgsino mass parameter $\mu$; the third generation slepton mass parameters $M_{\widetilde{L}_3}$ and $M_{\widetilde{E}_3}$; the third generation squark mass parameters $M_{\widetilde{Q}_3}$, $M_{\widetilde{U}_3}$, and $M_{\widetilde{D}_3}$; the third generation the trilinear couplings $A_{t,b,\tau}$, are separately determined for each scenario.  However, in the first three scenarios, the parameter $X_{t}=A_{t}-\mu \cot\beta$ is set as an input parameter instead of the parameter $A_{t}$. In this study, for the \textit{CP}-conserving MSSM, all SUSY input parameters are chosen to be real and positive. We give a list of the input parameters for each scenario in Table~\ref{tab:scenarios}.
\begin{table}[htb]
\caption{The input parameters for each scenario, in which all masses are given in TeV. The symbol ``*" means that $A_{t}$ is taken as $A_{t}=X_{t}+\mu \cot\beta$. In the last two rows, the values of $\tan\beta$ and $m_{A}$ are given for benchmark points corresponding to each scenario.}\label{tab:scenarios}
\begin{ruledtabular}
\begin{tabular}{lcccc}
 &\multicolumn{1}{c}{$m_h^{125}$}&\multicolumn{1}{c}{$m_h^{125}(light~\widetilde\tau)$}&\multicolumn{1}{c}{$m_h^{125}(light~\widetilde\chi)$}
&\multicolumn{1}{c}{$m_h^{125}(align)~$}\\
\hline
$M_{\widetilde{Q}_3,\widetilde{D}_3,\widetilde{U}_3}$  &1.5&1.5&1.5&2.5\\
$M_{\widetilde{L}_3,\widetilde{E}_3}$                  &2.0&~~~0.350&2.0&2.0\\
$\mu$                                                  &1.0&1.0 &~~~0.180        &7.5\\
$M_{1}$                                                &1.0&~~~0.180 &~~~0.160      &0.5\\
$M_{2}$                                                &1.0&~~~0.300 &~~~0.180       &1.0\\
$M_{3}$                                                &2.5&2.5&2.5 &2.5\\
$X_{t}$                                                &2.8&2.8&2.5  &$\cdot\cdot\cdot$\\
$A_{t}$                                                &*&*&*&6.25\\
$A_{b}$                                                &$A_{t}$&$A_{t}$&$A_{t}$&$A_{t}$\\
$A_{\tau}$                                             &$A_{t}$&~~~0.800&$A_{t}$       &$A_{t}$\\
\hline
                                                       &BP1&BP2&BP3&BP4\\
\hline
$\tan\beta$                                              &10&10&10 &10\\
$m_{A}$                                                  &1.5&1.5&1.5 &0.5\\
\end{tabular}
\end{ruledtabular}
\end{table}

In the $m_h^{125}$ scenario, all sparticles are relatively heavy; hence, they have only a mild effect on productions and decays of the Higgs bosons. Therefore, the phenomenology of this scenario is similar to that of a type-2 THDM with MSSM-inspired couplings of Higgs. The masses of the gluino and third-generation squarks are allowed by the available bounds from direct searches at the LHC. In the $m_h^{125}(light~\widetilde\tau)$ and $m_h^{125}(light~\widetilde\chi)$ scenarios, the colorless sparticles (staus and in one case, neutralinos and charginos) are relatively light, and the LSP is the lightest neutralino. The masses of the gluino, sbottom, and stop are the same as in the $m_h^{125}$ scenario, but the trilinear interaction term for the staus and the stop mixing parameter $X_t$ are reduced in the $m_h^{125}(light~\widetilde\tau)$ and $m_h^{125}(light~\widetilde\chi)$ scenarios, respectively. These two scenarios can be considerably effective on the Higgs phenomenology (see, e.g., Refs.~\cite{Carena,Carena2}) via loop contributions to the couplings of Higgs boson to particles of SM as well as via direct decays of the Higgs bosons into sparticles if kinematically possible~\cite{mh125}. At low values of $\tan\beta$, the $m_h^{125}(align)$ scenario is defined by alignment without decoupling. To obtain an acceptable prediction for $m_h$ as well as to achieve alignment without decoupling, in the $m_h^{125}(align)$ scenario, the parameters determining the stop masses are remarkably larger values than in the other scenarios. For each scenario,  taking both theory and experimental uncertainties into account, the impact on the parameter space of the available constraints from Higgs searches at the LHC, Tevatron, and LEP have been investigated in Ref.~\cite{mh125}. In this study, four benchmark points (BPs), which are compatible with the most recent results of the LHC for the bounds on masses and couplings of new particles, and the Higgs-boson properties are chosen.
These are checked by the \textsc{HiggsBounds}~\cite{HBounds} and \textsc{HiggsSignals}~\cite{HSignals} with results of 86 analyses.

Moreover, the SM input parameters are fixed as $m_h= 125.09\gev$, $m_W= 80.385\gev$, $m_Z= 91.1876\gev$,  $m_t^{pole}=172.5\gev$, $m_b(m_b)=4.18\gev$, and $\alpha^{-1}= 137.036$~\cite{PDG}. 
For all scenarios, the masses of Higgs bosons are computed to two-loop accuracy with help of the \textsc{FeynHiggs}. The theoretical uncertainty in the prediction of \textsc{FeynHiggs} is estimated as $\Delta m_h^{theory}=\pm3\gev$ for the Higgs masses~\cite{Degrassi,Allanach}. The dependence of properties of the Higgs boson on other lepton and quark masses is not very pronounced, and the default values of \textsc{FeynHiggs} are considered. The values of the input flags of \textsc{FeynHiggs} are set such that the evaluation includes full next-to-leading logarithms (NLL) and partial next-to-NLL resummation of the logarithmic corrections.

\section{Numerical results and discussion}\label{sec:results}
In this section, the numerical predictions of the single production of the neutral Higgs bosons in association with a photon in electron-positron collisions are discussed in detail for both the SM and MSSM, focusing on individual contributions from each type of one-loop diagram and polarizations of initial beams. For each benchmark scenario given in Table~\ref{tab:scenarios}, the numerical evaluation of the total cross sections and individual contributions of one-loop amplitudes for the process $e^{+} e^{-} \to h\gamma$ has been carried out as a function of the center-of-mass energy. Furthermore, the total cross section of $e^- e^+ \rightarrow \gamma h^0$ and $e^- e^+ \rightarrow \gamma A^0$ are scanned over the plane $m_A-\tan\beta$.

\subsection{Process $e^- e^+ \to \gamma h^0$ in the SM}
\begin{figure}[!htb]
    \begin{center}
\includegraphics[scale=0.43]{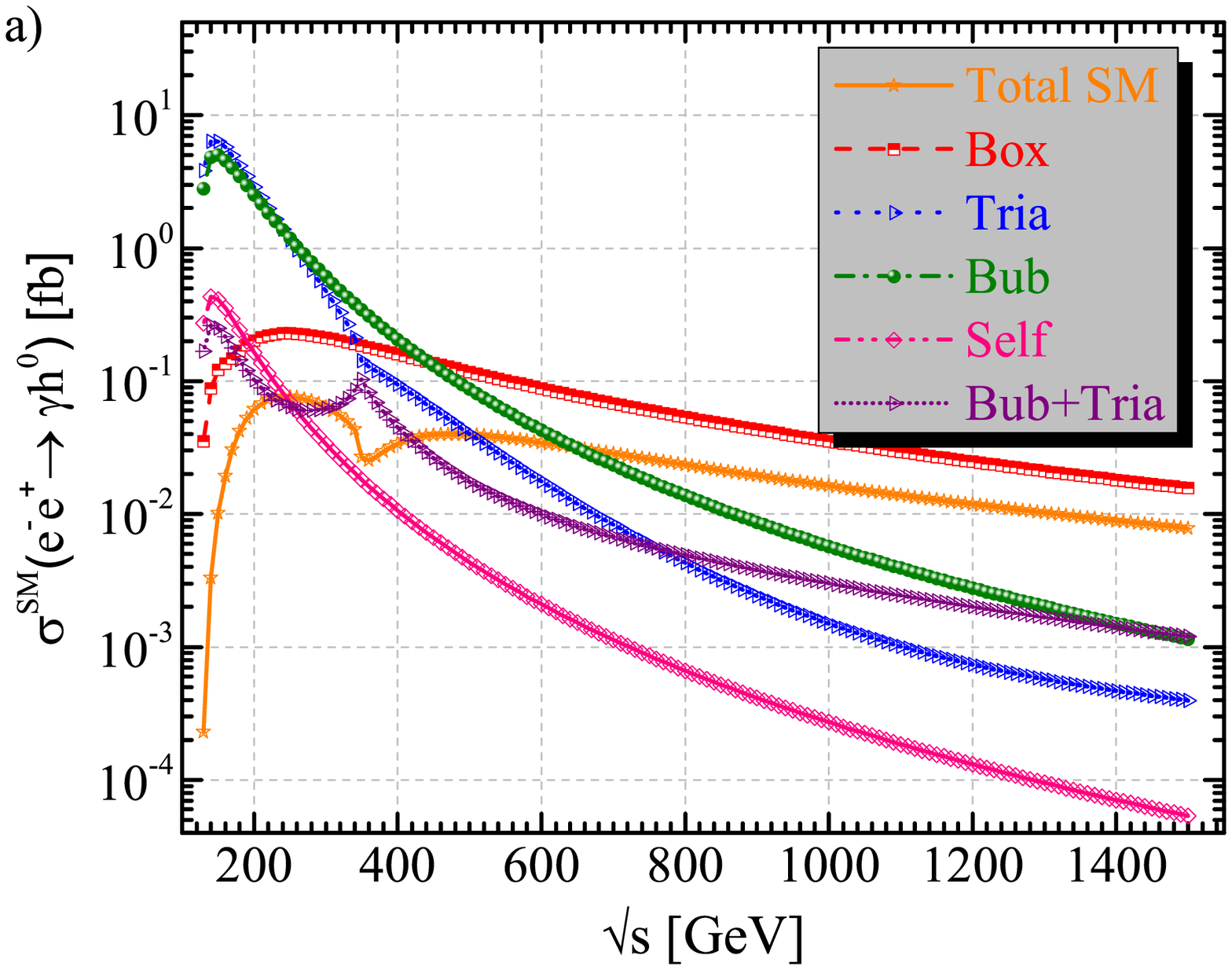}
\includegraphics[scale=0.43]{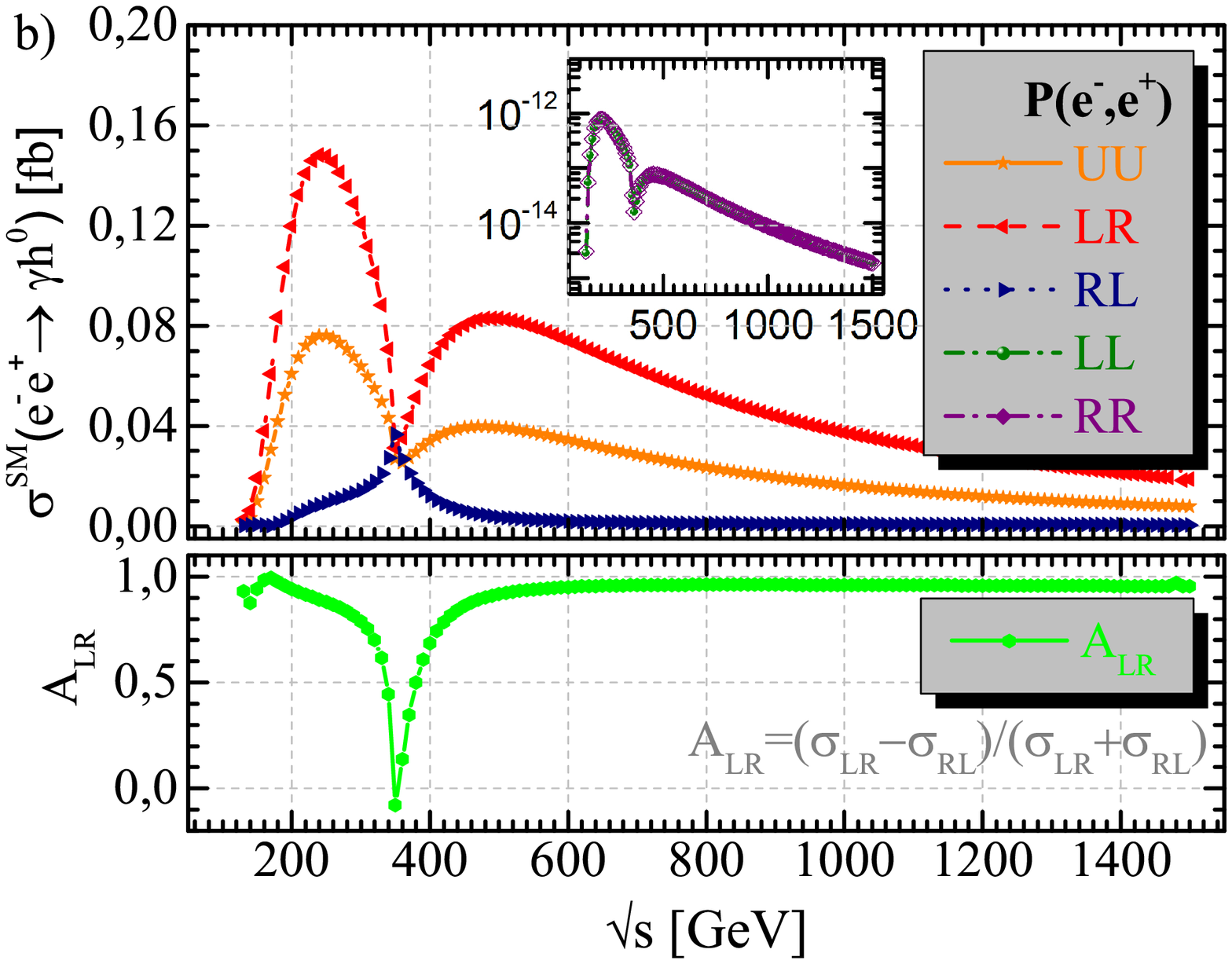}
     \end{center}
     \vspace{ -4mm}
\caption{(color online). The cross sections of process $e^- e^+ \to \gamma h^0$ in the SM as a function of center of mass energy for a) the individual amplitude from each type of one-loop diagram and b) various polarization cases of initial beams. Also, the left-right asymmetry $A_{LR}$ is shown at the lower panel of part (b).}
\label{fig:SM}
\end{figure}

In Fig.~\ref{fig:SM}, the SM cross section of $e^{+} e^{-} \to h\gamma$ is given as a function of centre-of-mass energy in the range from 100 to 1500 GeV, focusing on individual one-loop contributions, and polarizations of initial beams.
\begin{figure*}[!t]
    \begin{center}
\includegraphics[scale=0.43]{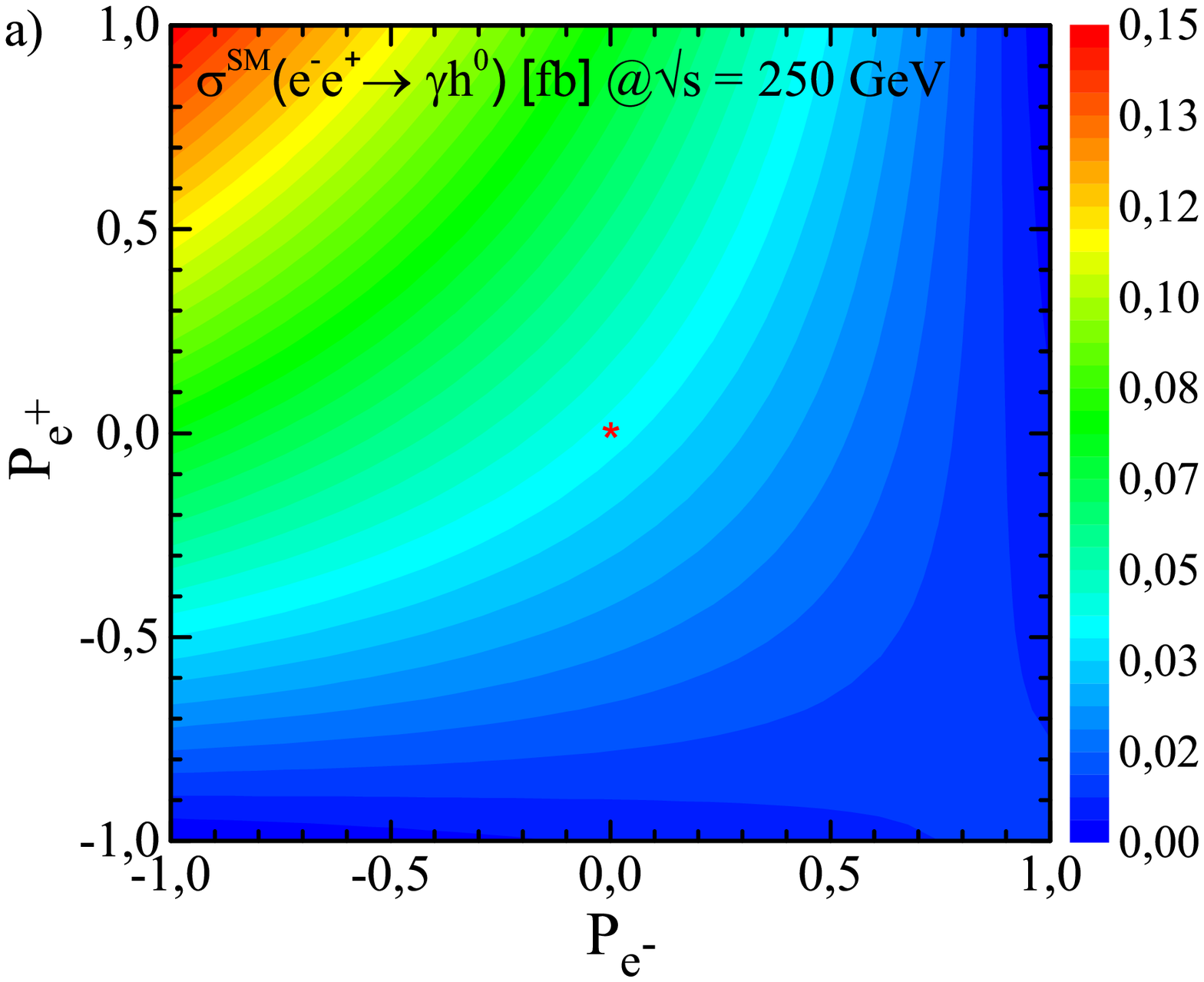}
\includegraphics[scale=0.43]{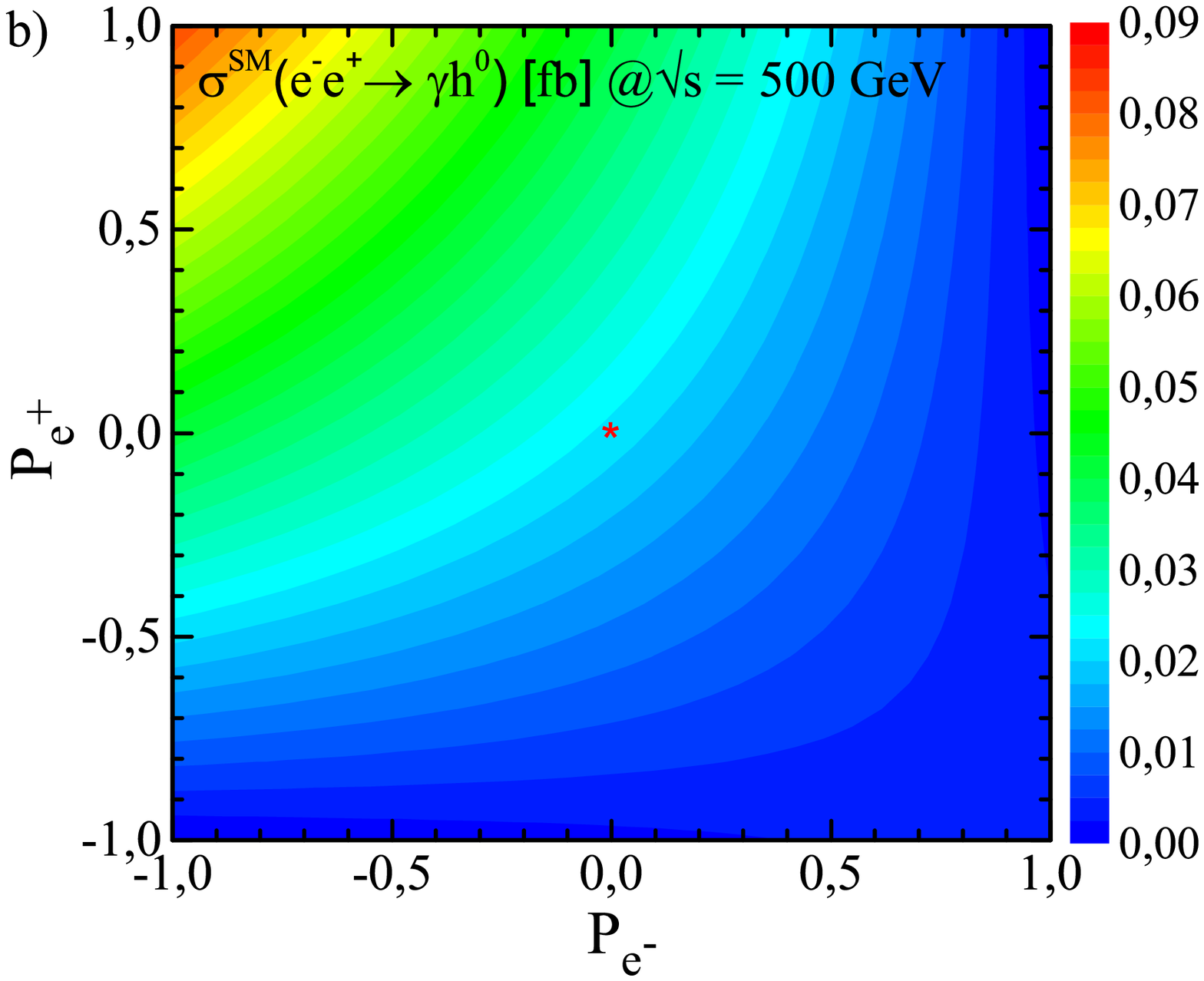}
     \end{center}
     \vspace{ -4mm}
\caption{(color online). The cross section of process $e^- e^+ \to \gamma h^0$ in the SM as a two-dimensional function of $P_{ e^-}$ and $P_{ e^+}$ at a) $\sqrt{s}=250\gev$ and b) $\sqrt{s}=500\gev$, where the color heat map corresponds to the total cross section (in fb) in the scan region. The red stars denote the unpolarized cross section at point $(P_{ e^-}, P_{e^+}) =(0,0)$.}
\label{fig:SM2D}
\end{figure*}
It is seen that the total cross section increases quickly with the opening of the phase space and then decreases near
$\sqrt{s}\sim 2\times m_t$ (two times mass of the top quark) close to the $t \bar t$ threshold, with increments of $\sqrt{s}$. As a function of $\sqrt{s}$, a maximum of 0.076 fb is reached around $\sqrt{s}=250\gev$. It is seen that the cross section is very sensitive to the magnitudes of each amplitude and the relative phases between them. At high center-of-mass energy, at which the self-energy\footnote{The main self-energy contributions come from diagrams in which the virtual $Z$ line with $Z-\gamma$ mixes through one-loop.}, the triangle-, and the bubble-type contributions are suppressed, $\sigma^{SM}(e^{+} e^{-} \to h\gamma)$ is dominated by the box-type contributions. However, the bubble-type contribution is larger than the triangle-type contribution, and their interference (bub+tri) makes a much smaller contribution from each of them in the region of $\sqrt{s}\leq700\gev$, since they nearly destroy each other. The (bub+tri) contribution is enhanced by the threshold effect when $\sqrt{s}$ is close to $2\times m_t$, since the $t \bar t$ threshold is seen at this energy. The top-quark and $W$-boson contributions make destructive interference, and the top contribution is maximal near the $t \bar t$ threshold. After passing the threshold of  $t \bar t$, the cross section scales like $1/s$ and hence drops steeply. Combining all the contributions, the total cross section ultimately has the first peak near $\sqrt{s}=250\gev$ and the second one near $\sqrt{s}=500\gev$. Note that the total cross section with a completely polarized left-handed electron $e^{-}_L$ and right-handed positron $e^{+}_R$, $\sigma^{SM}(e^{-}_L e^{+}_R \to h\gamma)$ can be enhanced by about a factor between $2$ and $4$, compared with the unpolarized case. The longitudinal polarization of both the electron and positron beams is therefore significant to enhance the cross section. At $\sqrt{s}=250\gev$,  $\sigma^{SM}(e^{-}_L e^{+}_R \to h\gamma)$ reaches a value of $0.15$ fb. However, as expected, the cross sections for polarization cases of $e^{-}_L e^{+}_L$ and $e^{-}_R e^{+}_R$ are very small [see the insert figure in Fig.~\ref{fig:SM}(b)]. The left-right asymmetry $A_{LR}$  has a peak at the region of $\sqrt{s}\leq340\gev$. After passing the $t \bar t$ threshold, it remains nearly constant with a value of $0.96$.

In Figs.~\ref{fig:SM2D}(a) and~\ref{fig:SM2D}(b), the cross section of process $e^- e^+ \to \gamma h^0$ in the SM is also presented in the plane of $P_{ e^-}$ and $P_{ e^+}$ by varying from $-1$ to $+1$. Especially, the cross section reaches its sizable values in the regions of $0<P_{ e^+}\leq +1$ and $-1\leq P_{ e^-}<0$.  The polarized cross section is maximum at point $(P_{ e^-}, P_{e^+}) =(-1,+1)$, namely, a completely polarized left-handed electron and completely right-handed positron. The enhancement is raised up to a factor of 2 at the left top corner, compared with the unpolarized case. At $\sqrt{s}=250\gev$, the polarized cross section $\sigma^{SM}(P_{ e^-},P_{ e^+})$ reaches up to $\sigma^{SM}(-0.8,+0.3) = 0.09$ fb and $\sigma^{SM}(-0.8, +0.6) = 0.11$ fb. At $\sqrt{s}=500\gev$, the polarized cross section $\sigma^{SM}(P_{ e^-},P_{ e^+})$ reaches up to $\sigma^{SM}(-0.8,+0.3) = 0.05$ fb and $\sigma^{SM}(-0.8, +0.6) = 0.06$ fb.

\subsection{Process $e^- e^+ \to \gamma h^0$ in the MSSM}
\begin{figure}[!ht]
    \begin{center}
\includegraphics[scale=0.45]{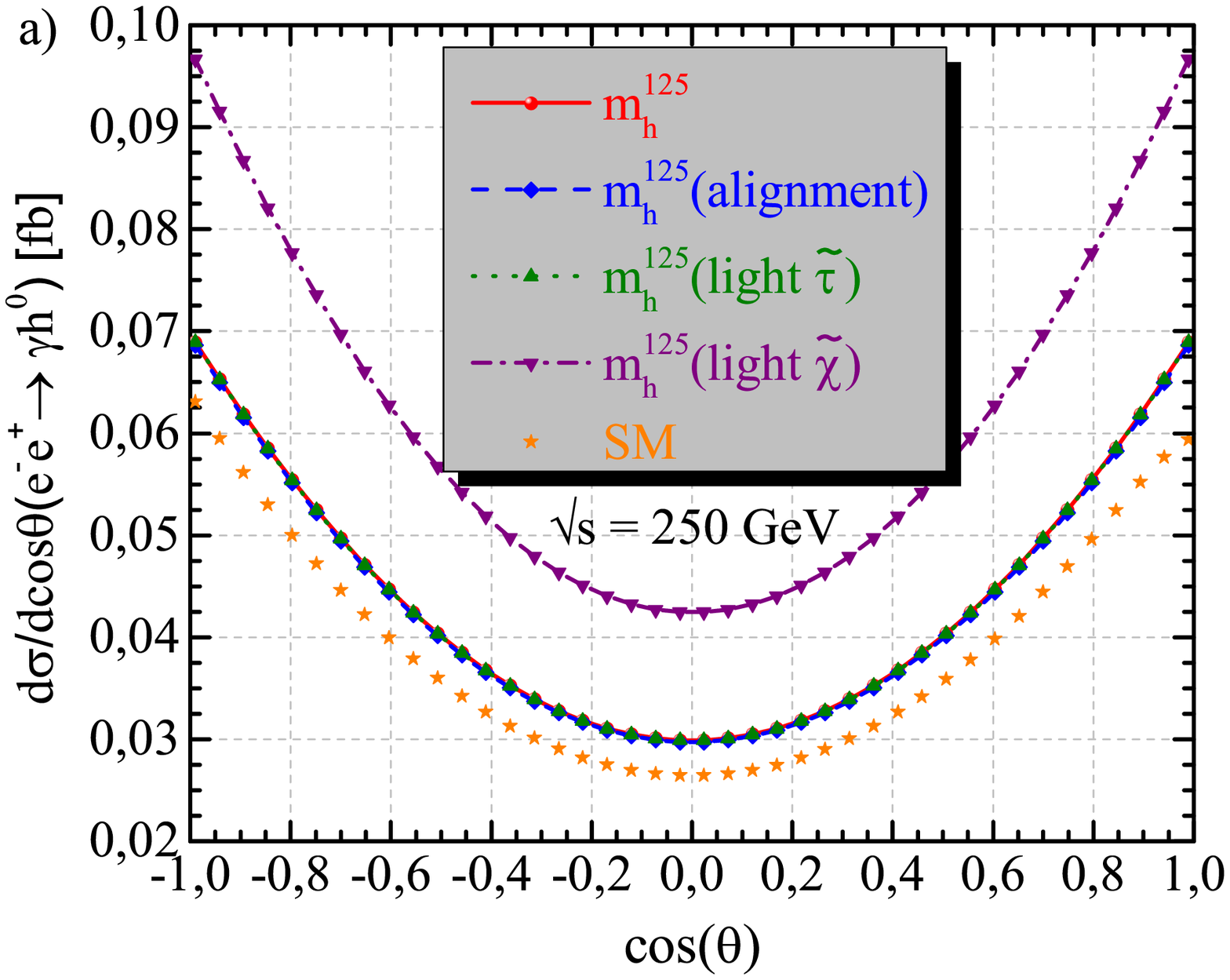}
\includegraphics[scale=0.434]{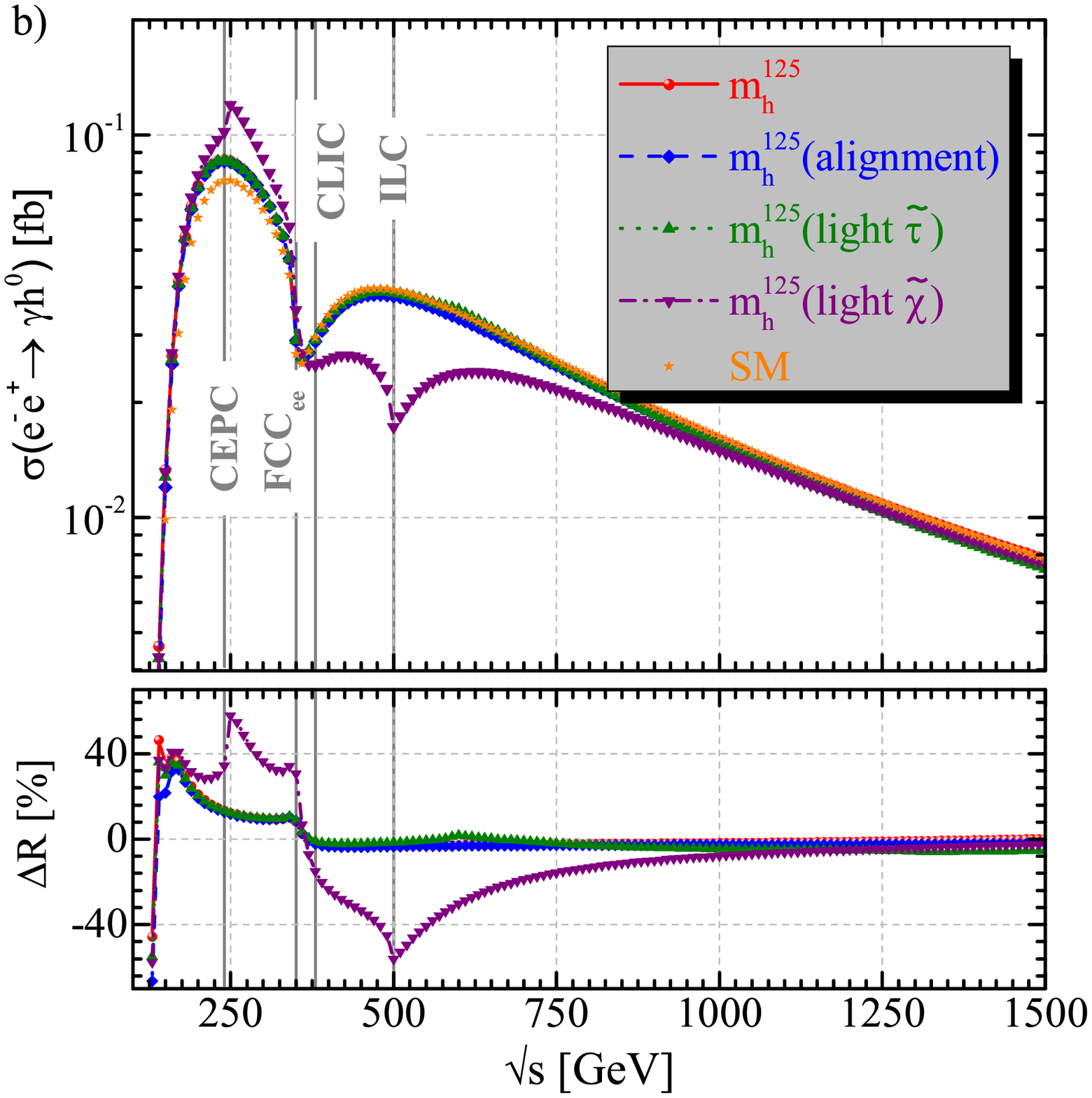}
     \end{center}
\vspace{ -5mm}
\caption{(color online). a) Differential cross section at $\sqrt{s}=250\gev$ and b) total cross section of process $e^- e^+ \to \gamma h^0$ as a function of center of mass energy for each benchmark point defined in the scenarios of $m_h^{125}$, $m_h^{125}(alignment)$, $m_h^{125}(light~\widetilde\tau)$, and $m_h^{125}(light~\widetilde\chi)$. The vertical solid lines indicate to the proposed energy of each future lepton collider. The lower panel in part (b) shows the deviations from the predictions of SM.}
\label{fig:mh125}
\end{figure}
In Fig.~\ref{fig:mh125}(a), the corresponding angular distribution $d\sigma/d \cos\theta$ at $\sqrt{s}=250\gev$ is exhibited for each benchmark point scenario given in Table~\ref{tab:scenarios} and the SM. It is clear that all distributions are rather symmetric and the distributions for benchmarks point scenarios in the MSSM are larger than those of the SM. In Fig.~\ref{fig:mh125}(b), the total cross section of process $e^- e^+ \to \gamma h^0$ in the MSSM is presented as a function of $\sqrt{s}$ from 100 GeV to 1.5 TeV for each benchmark point scenario. Additionally, to see deviations from the predictions of the SM, the ratio of the total cross sections
\begin{equation}  \label{eq:delta}
\Delta \text{R}=\frac{(\sigma_{\text{MSSM}}-\sigma_{\text{SM}})}{\sigma_{\text{SM}}}\times100
\end{equation}
is evaluated for each scenario, and this is presented in the lower panel of Fig.~\ref{fig:mh125}(b). The distribution of the cross section in the MSSM has the same trend as the SM distribution. Because of the $t \bar t$ threshold, a dip appears at $\sqrt{s}\approx340\gev$ for all scenarios. However, the next dip at $\sqrt{s}\approx 500\gev$ is observed for only the $m_h^{125}(light~\widetilde\chi)$ scenario due to the threshold $m_{\chi^\pm_2}+m_{\chi^\pm_2}=\sqrt{s}$. When the $\sqrt{s}$ runs from $250$ to $500\gev$, the total cross section decreases from 0.085 to 0.038 fb in the $m_h^{125}$, $m_h^{125}(light~\widetilde\tau)$ and $m_h^{125}(alignment)$ scenarios, while it decreases from 0.12 to 0.017 fb in the $m_h^{125}(light~\widetilde\chi)$ scenario. The remarkable deviations from the predictions of the SM are seen in the $m_h^{125}(light~\widetilde\chi)$ scenario such that at $\sqrt{s}=250\gev$ the production rate is enhanced by 58$\%$, whereas at $\sqrt{s}=500\gev$, it is reduced by 56$\%$. In other scenarios, i.e., $m_h^{125}$, $m_h^{125}(light~\widetilde\tau)$, and $m_h^{125}(alignment)$, however, there is a deviation of about 10$\%$ from the predictions of the SM at $\sqrt{s}=250\gev$. For the scenarios in which all sparticles are too heavy to be produced directly at the selected center of mass energy, the MSSM contributions are small and will, therefore, be difficult to detect. For the $m_h^{125}(light~\widetilde\chi)$ scenario, the unpolarized cross section is around $0.10$ fb, $0.035$ fb, $0.025$ fb, and  $0.017$ fb for the planned CEPC (at $\sqrt{s}=240\gev$), FCC$_{\rm ee}$ (at $\sqrt{s}=350\gev$), CLIC (at $\sqrt{s}=380\gev$), and ILC (at $\sqrt{s}=500\gev$) projects, respectively. For other scenarios, the unpolarized cross section reaches values of $0.086$ fb, $0.029$ fb, $0.029$ fb, and  $0.038$ fb for the planned CEPC, FCC$_{\rm ee}$, CLIC-380, and ILC-500 projects, respectively. The energy-dependent structure of the cross section appears at a value of $\sqrt{s}$, which is close to two times mass of some particles, i.e., threshold effects.

\begin{figure}[!t]
    \begin{center}
\includegraphics[scale=0.42]{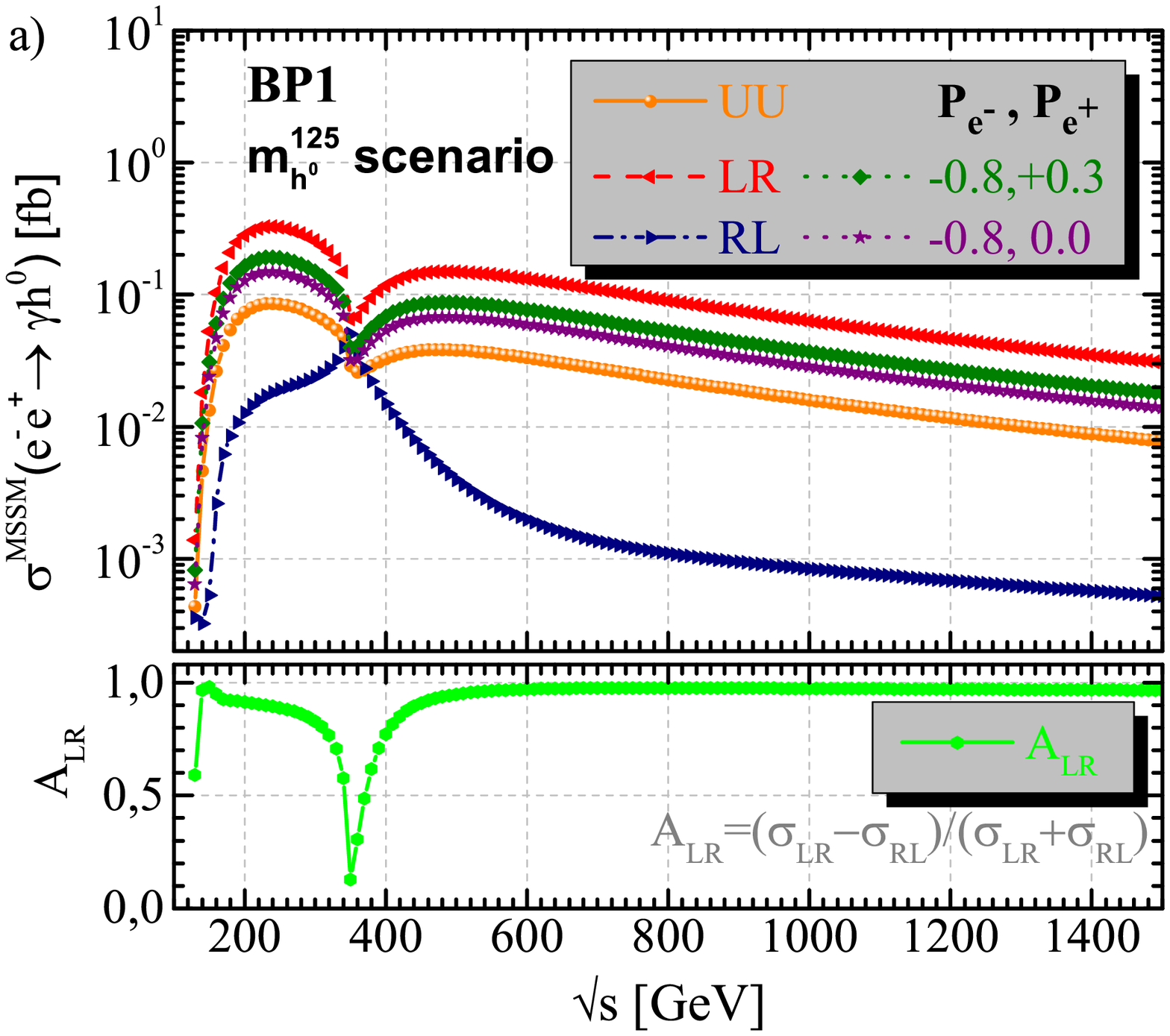}
\includegraphics[scale=0.42]{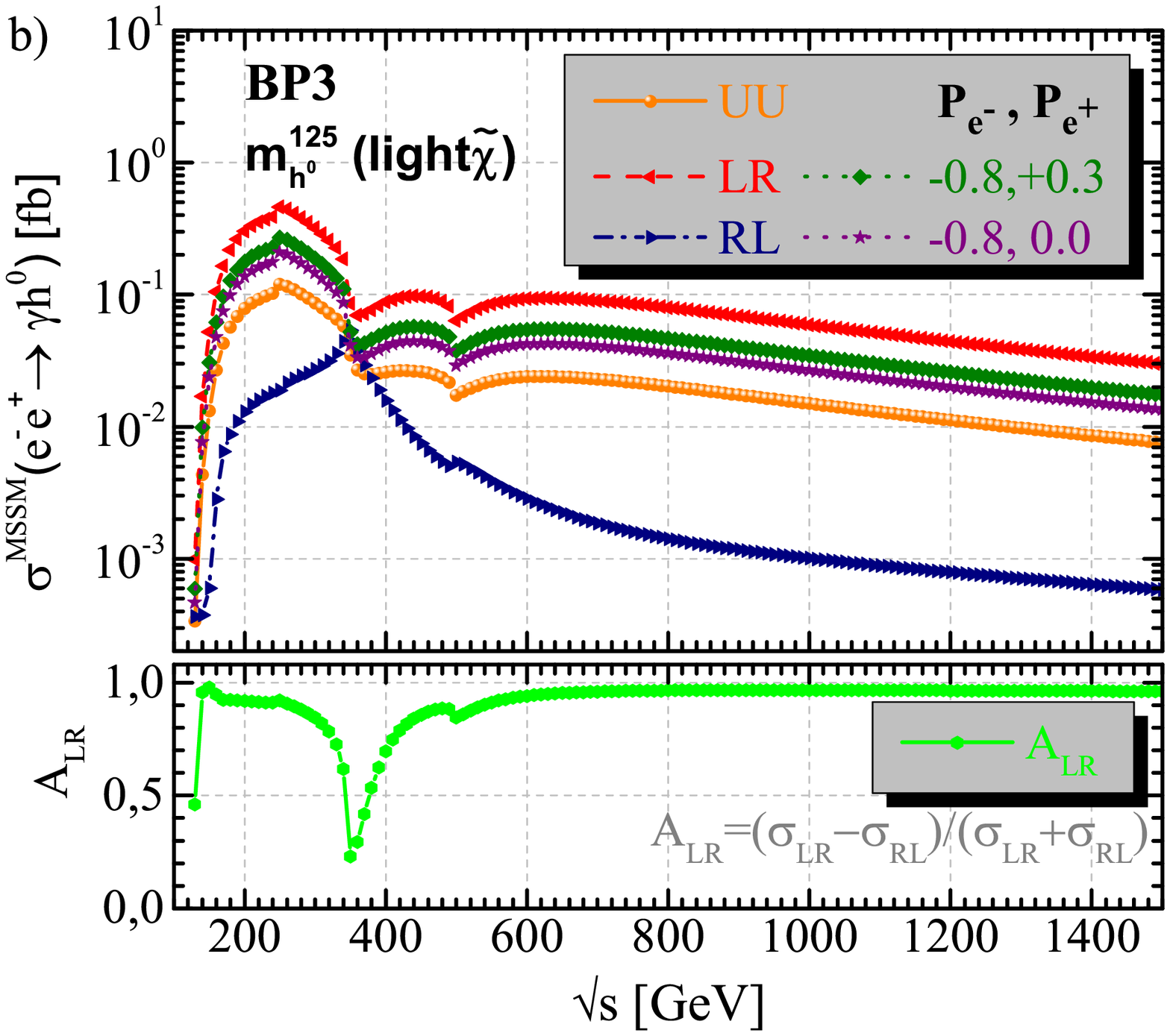}
     \end{center}
\vspace{ -5mm}
\caption{(color online). The polarized cross section of process $e^- e^+ \to \gamma h^0$ for various polarizations of the initial beams as a function of center of mass energy for two benchmark points defined in the scenarios of a) $m_h^{125}$ and b) $m_h^{125}(light~\widetilde\chi)$. Also, the left-right asymmetry $A_{LR}$ is shown in the lower panel of each part.}
\label{fig:MSSMpol}
\end{figure}

In Fig.~\ref{fig:MSSMpol}, the polarized cross sections of $e^{+} e^{-} \to h\gamma$ are given as a function centre-of-mass energy in the range from 100 to 1500 GeV, for two benchmark points defined in the scenarios $m_h^{125}$ and $m_h^{125}(light~\widetilde\chi)$. Note that for the other two scenarios, distributions of the polarized cross section are not shown here because they are similar to that of the $m_h^{125}$ scenario.
The total cross section with a completely polarized left-handed electron and right-handed positron, $\sigma^{\text{MSSM}}(e^{-}_L e^{+}_R \to h\gamma)$ for each benchmark point, can be enhanced by about a factor between $3$ and $4$, compared with the unpolarized case. The longitudinal polarization of both the positron and electron beams is, hence, significant to enhance the cross section. At $\sqrt{s}=250\gev$,  $\sigma^{\text{MSSM}}(e^{-}_L e^{+}_R \to h\gamma)$ reaches values of $0.32$ fb and $0.46$ fb for the benchmark points scenarios $m_h^{125}$ and $m_h^{125}(light~\widetilde\chi)$, respectively.
For BP1($m_h^{125}$), the polarized cross section $\sigma^{\text{MSSM}}(P_{ e^-},P_{ e^+})$ reaches up to $\sigma^{\text{MSSM}}(-0.8,+0.3) = 0.19$ fb and $\sigma^{\text{MSSM}}(-0.8, 0.0) = 0.15$ fb at $\sqrt{s}=250\gev$. On the other hand, for BP3($m_h^{125}(\widetilde\chi)$), the polarized cross section $\sigma^{\text{MSSM}}(P_{ e^-},P_{ e^+})$ reaches up to $\sigma^{\text{MSSM}}(-0.8,+0.3) = 0.27$ fb and $\sigma^{\text{MSSM}}(-0.8, 0.0) = 0.21$ fb. The left-right asymmetry $A_{LR}$ has a peak at the region of $\sqrt{s}\leq340\gev$. After passing the $t \bar t$ threshold, it remains nearly constant with a value of $0.97$. Furthermore, the cross sections are sorted according to various polarizations of initial beams as follows: $\sigma$(LR)$>\sigma$(-0.8,+0.3)$>\sigma$(-0.8,0.0)$>\sigma$(UU)$>\sigma$(RL) for each benchmark point. It is seen that the basic size of the total cross sections is enhanced by a factor of 4, depending on the polarizations of the initial $e^-$ and $e^+$ beams.

\begin{figure*}[!ht]
    \begin{center}
\includegraphics[scale=0.41]{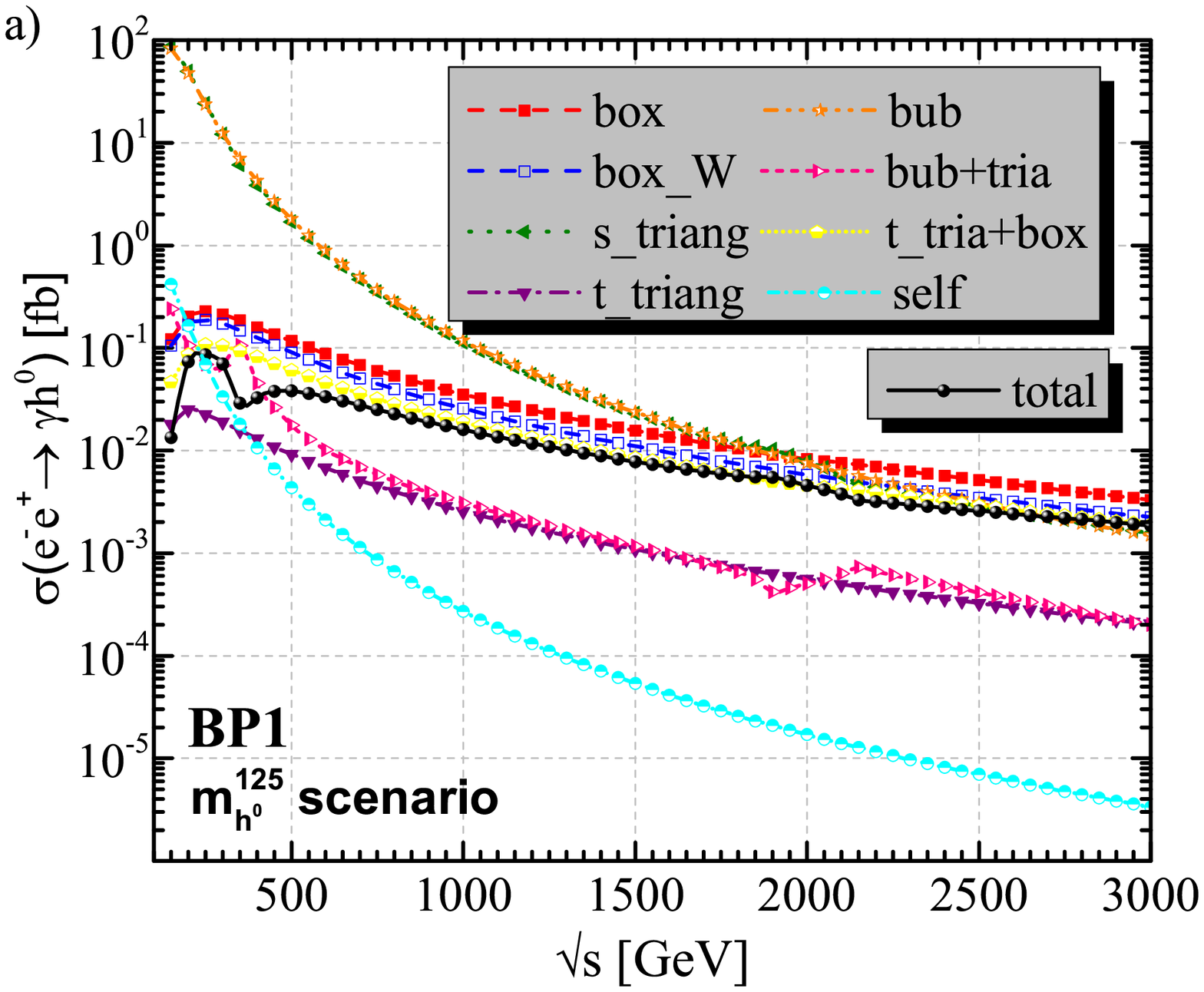}
\includegraphics[scale=0.41]{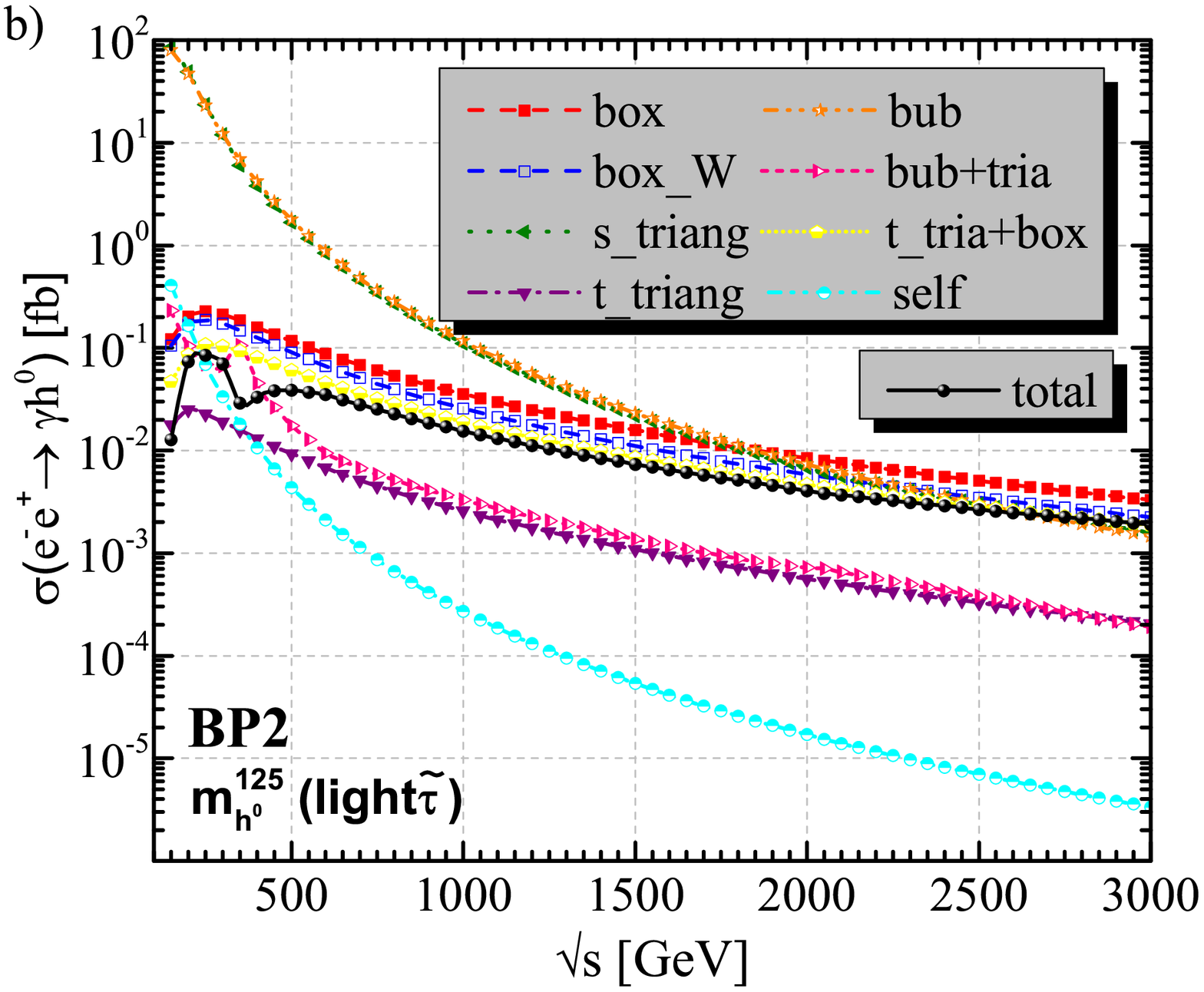}
\includegraphics[scale=0.41]{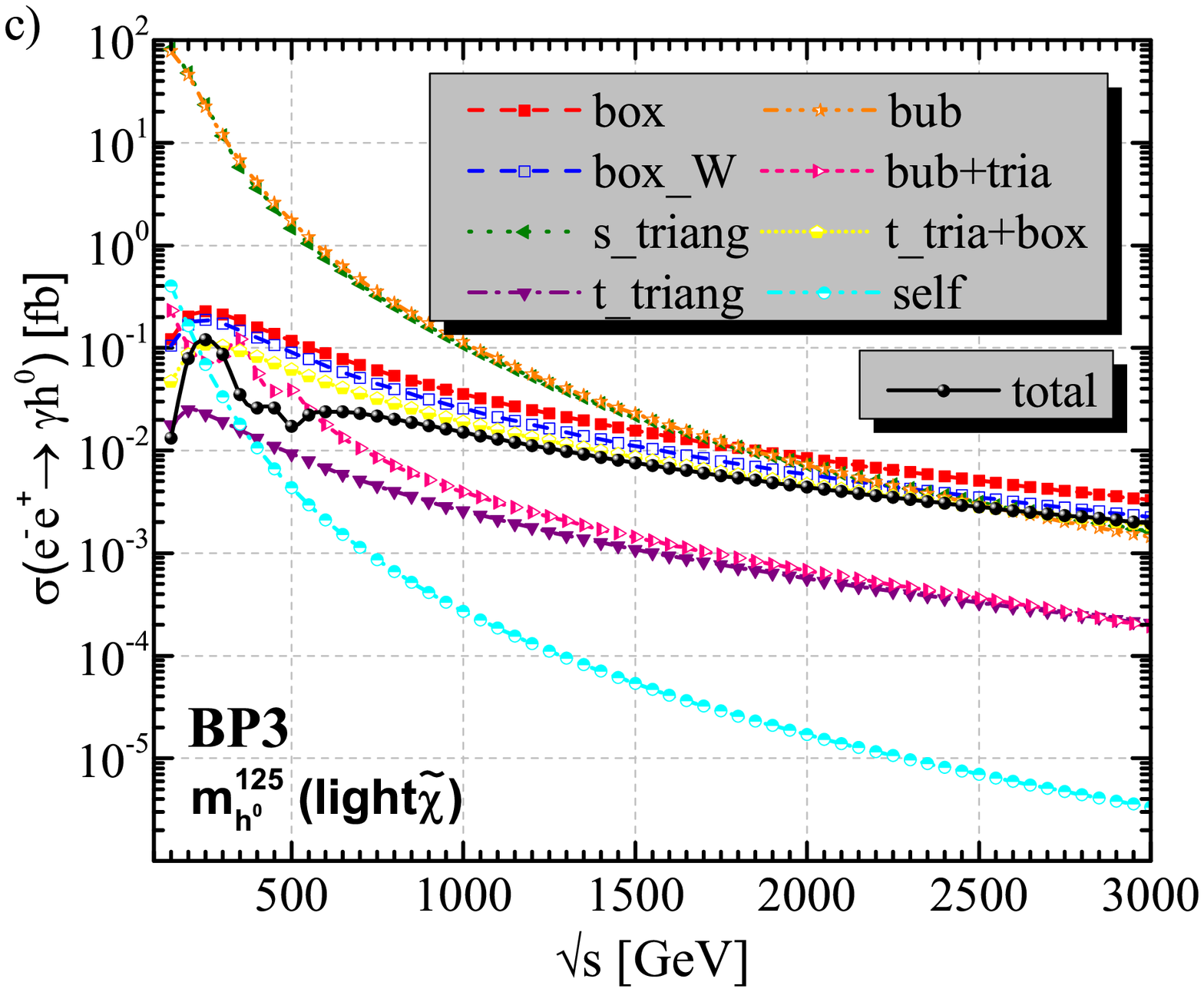}
\includegraphics[scale=0.41]{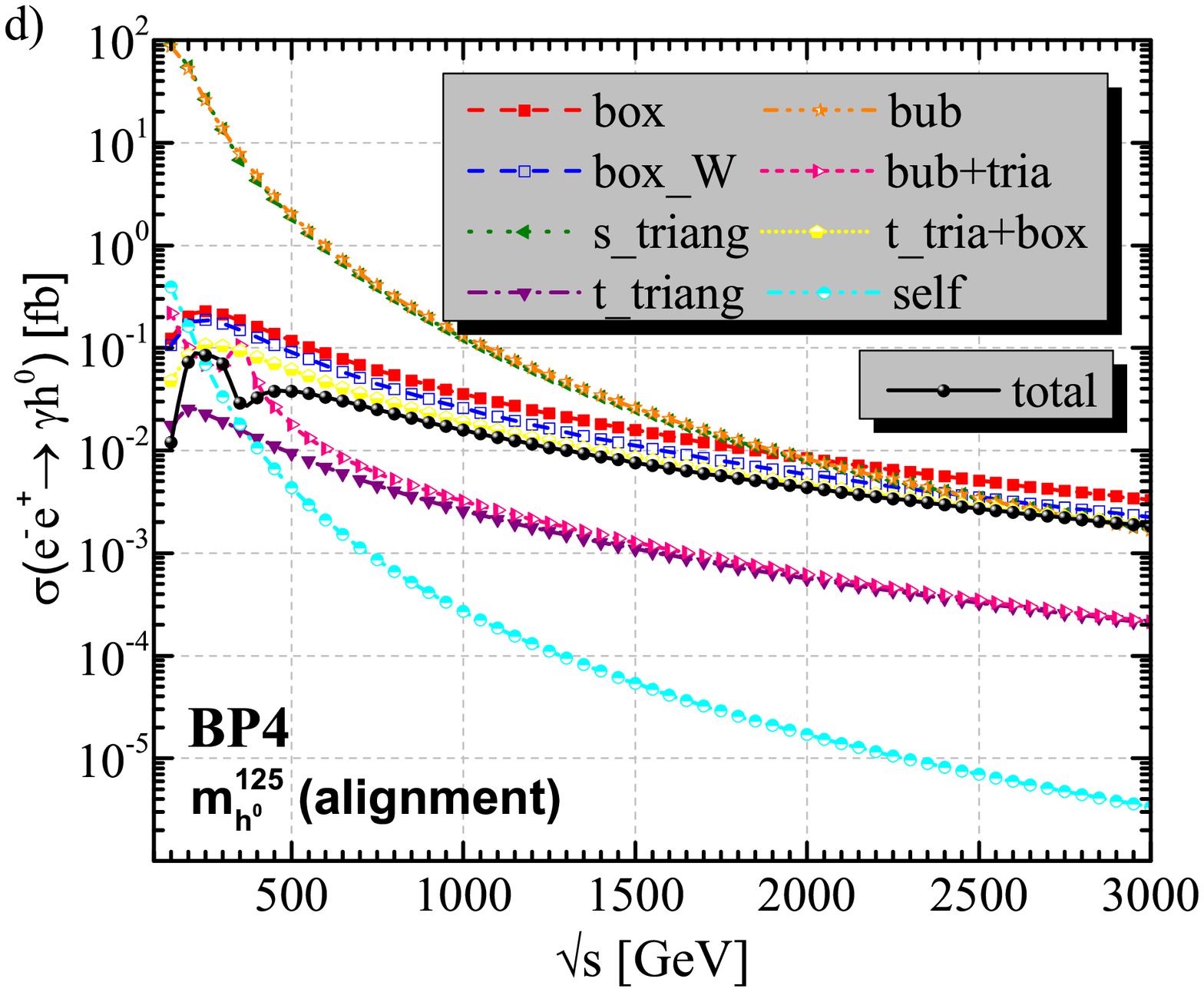}
     \end{center}
\vspace{ -5mm}
\caption{(color online). The individual contributions from each type of diagram to the total cross section of process $e^- e^+ \to \gamma h^0$ as a function of center of mass energy for benchmark points defined in the scenarios of a) $m_h^{125}$, b) $m_h^{125}(light~\widetilde\tau)$, c) $m_h^{125}(light~\widetilde\chi)$, and d) $m_h^{125}(alignment)$.}
\label{fig:MSSMindiv}
\end{figure*}
Figure~\ref{fig:MSSMindiv} shows the effect of the individual contributions coming from different types of diagrams on the total cross section of process $e^- e^+ \to \gamma h^0$ as a function of the center-of-mass energy ranging from 100 GeV to 3 TeV for each benchmark point scenario. Here, the abbreviations  ``box'', ``bub'',``qua'',``s\_triang'' ,``t\_triang'', ``self'' and ``all'' indicate the contributions of box-type (diagrams b$_{1-12}$ in Fig.~\ref{fig:box}), bubble-type (diagrams q$_{1-7}$ in Fig.~\ref{fig:qua}), quartic-type (diagram q$_{8}$ in Fig.~\ref{fig:qua}), $s$-channel triangle-type (diagrams t$_{1-7}$ in Fig.~\ref{fig:tria}), $t$- and $u$-channel triangle-type (diagrams t$_{8-18}$ in Fig.~\ref{fig:tria}), and self-energy (diagrams s$_{1-10}$ in Fig.~\ref{fig:self}) diagrams and all of Feynman diagrams, respectively.
``t\_tria+box'' denotes the total contribution from all $t$- and $u$-channel diagrams.
``box\_W'' represents contributions from the box-type diagrams with one or more $W$ bosons (diagrams b$_{3}$, b$_{4}$, b$_{7}$, and b$_{8}$ in Fig.~\ref{fig:box}). Also, ``bub+tria'' denotes the contribution from interference between bubble-type and triangle-type diagrams.

It is clearly seen that the cross section is very sensitive to the magnitudes of each amplitude and the relative phases between them in all of the considered scenarios. $\sigma^{\text{MSSM}}(e^{+} e^{-} \to h\gamma)$ is dominated by the s\_triangle- and the bubble-type contributions at low center-of-mass energy. On the other hand, at high energies, $\sqrt{s} \geq 2\tev$, these contributions are suppressed as 1/s, and hence the box-type contributions become greater than them. However, the triangle- and bubble-type contributions are almost equal, and the interference among them (bub+tria) makes a much smaller contribution than any of them (by 2 orders of magnitude) because they make a destructive interference. On the other hand, the box-type contribution is larger by 1 order of magnitude than that of this interference. The $t$-channel contributions come from box-type and t\_triang-type diagrams; however, the contribution of box-type diagrams is reduced by t\_triang diagrams. This means that there is some cancellation between t\_triang-type diagrams and the box-type diagrams, as already mentioned. The total contribution from $t$-channels is smaller than $s$-channel contributions (buble-type and s\_triangle-type).

The dominant contribution of scalar fermions to the process comes from the $\tilde{\tau}$ loop. This can be figured out from the leading part of the amplitudes of the sfermions loop, which is proportional to $(g_a A_{\tilde{f}} +g_b \mu \tan\beta) \sin2\theta_{\tilde{f}}/m^2_{\tilde{f}}$~\cite{Djouadi2}. A large mixing parameter $A_t$ or heavy stops are needed in order to satisfy the requirement of the Higgs mass. The light $\tilde{t}_1$ can be obtained by the large $A_t$ but accompanies a heavy $\tilde{t}_2$. This will lead to a small stop mixing angle and reduce the contribution from $\tilde{t}$-loops.
Consequently, at scenarios with the light stau, the triangle-type diagrams are dominated by the $\tilde{\tau}$ loop because of the $A_{\tau}$ term being large.

The box\_W contribution ends up dominating over all the other ones because the masses of the sfermion and charged Higgs boson are fixed at the TeV scale. About $70\%$ of the total contribution of box-type diagrams comes from box-type diagrams b$_{3}$, b$_{4}$, b$_{7}$, and b$_{8}$ with one or more $W$-bosons in the loop. Therefore, it is also possible to assess the contribution of box-type diagrams in terms of a single coupling (the Higgs-$W$-$W$ coupling) given in Eq.~\eqref{eq:lambda9}, which is proportional to the term $\sin({\beta-\alpha})$.

Table~\ref{tab:indivcont} shows the numerical results over the scenarios for center-of-mass energies of the planned CEPC (at $\sqrt{s}=250\gev$), FCC$_{\rm ee}$ (at $\sqrt{s}=350\gev$), and ILC (at $\sqrt{s}=500\gev$) projects. Overall, the $s$-channel triangle- and bubble-type diagrams make a dominant contribution to the total cross section in all scenarios. Therefore, the s\_triangle- and the bubble-type diagrams have a remarkable impact on production rate. Particularly, the total cross section of the production of $e^- e^+ \to \gamma h^0$ reaches a value of 0.12 fb at $\sqrt{s}= 250$ GeV and is more observable compared to others at the CEPC or ILC-250 for BP3. Additionally, the cross sections are sorted according to the BPs as $\sigma$(BP3)$>\sigma$(BP1)$\sim \sigma$(BP2)$\sim \sigma$(BP4) at low energies. Note that the basic size of the total cross sections is not very sensitive according to the BPs.
\begin{table}[hb]%
\caption{The individual contributions from each type of diagram to the total cross section of process $e^- e^+ \to \gamma h^0$ at the different center of mass energies for all BPs corresponding
to scenarios $m_h^{125}$, $m_h^{125}(light~\widetilde\tau)$, $m_h^{125}(light~\widetilde\chi)$, and $m_h^{125}(align)$.}\label{tab:indivcont}
\begin{tabular}{lclccccc}
 \hline \hline
BPs&$\sqrt{s}$~(GeV)&\textbf{Box}&\textbf{s\_tria}&\textbf{Bubble}&\textbf{t\_chan}&\textbf{Self}&\textbf{All}\\
 \hline
\multirow{3}*{BP1}&250&0.228&23.93&23.42&0.108&0.069&0.085\\
                          &350&0.185&~6.07&~6.98&0.093&0.018&0.029\\
                          &500&0.117&~1.70&~1.79&0.061&0.004&0.038\\
\hline
\multirow{3}*{BP2}&250&0.228&23.54&23.03&0.108&0.069&0.085\\
                                          &350&0.185&~5.98&~6.87&0.093&0.018&0.029\\
                                          &500&0.117&~1.67&~1.76&0.061&0.004&0.039\\
\hline
\multirow{3}*{BP3}&250&0.227&23.56&22.52&0.107&0.069&0.120\\
                                          &350&0.186&~5.78&~6.73&0.093&0.018&0.035\\
                                          &500&0.118&~1.46&~1.73&0.061&0.004&0.017\\
\hline
\multirow{3}*{BP4}&250&0.227&26.40&25.89&0.107&0.069&0.085\\
                                          &350&0.186&~6.79&~7.75&0.093&0.018&0.029\\
                                          &500&0.118&~1.89&~1.99&0.061&0.004&0.038\\
 \hline  \hline
\end{tabular}
\end{table}

\begin{figure*}[!ht]
    \begin{center}
\includegraphics[scale=0.44]{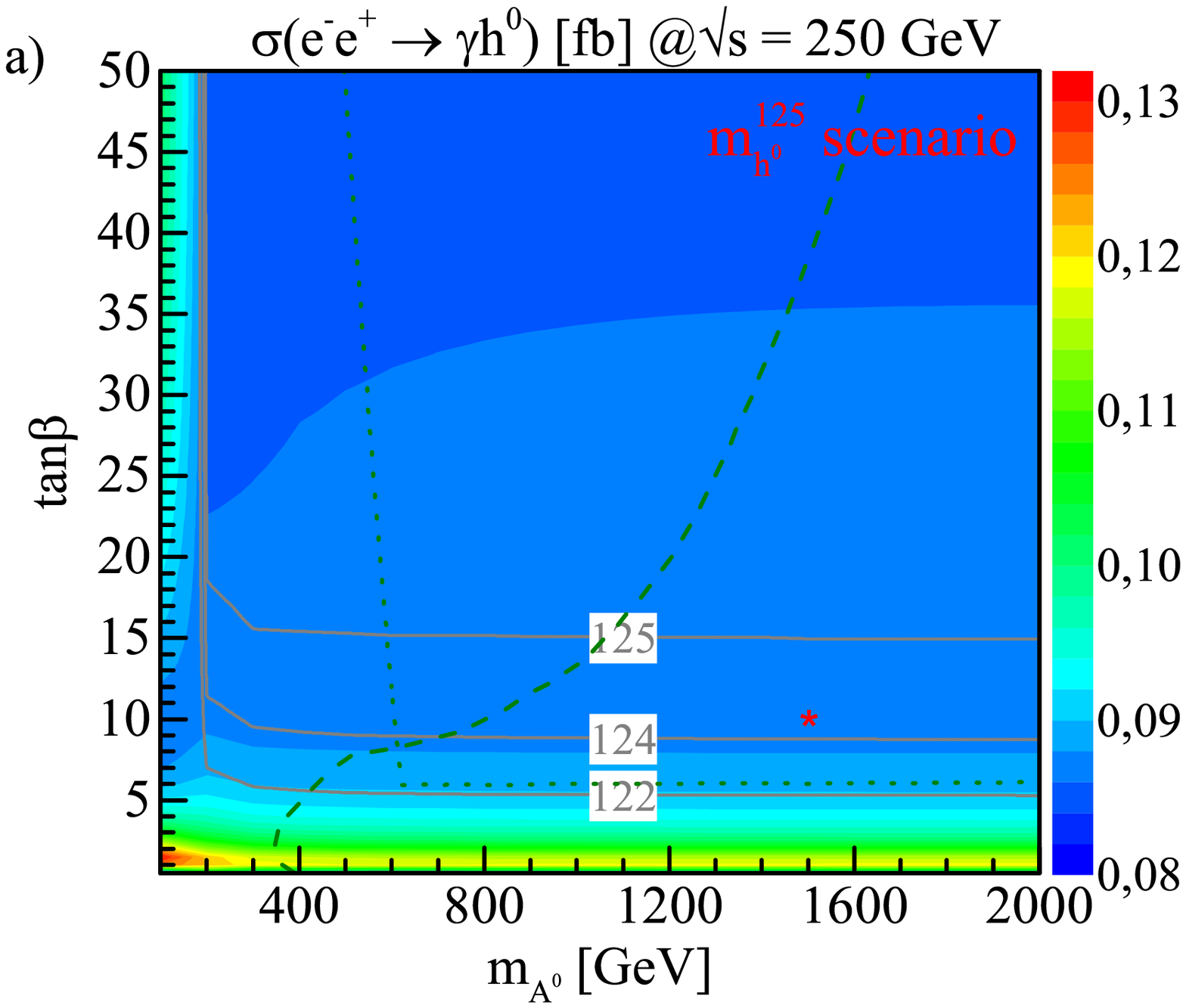}
\includegraphics[scale=0.44]{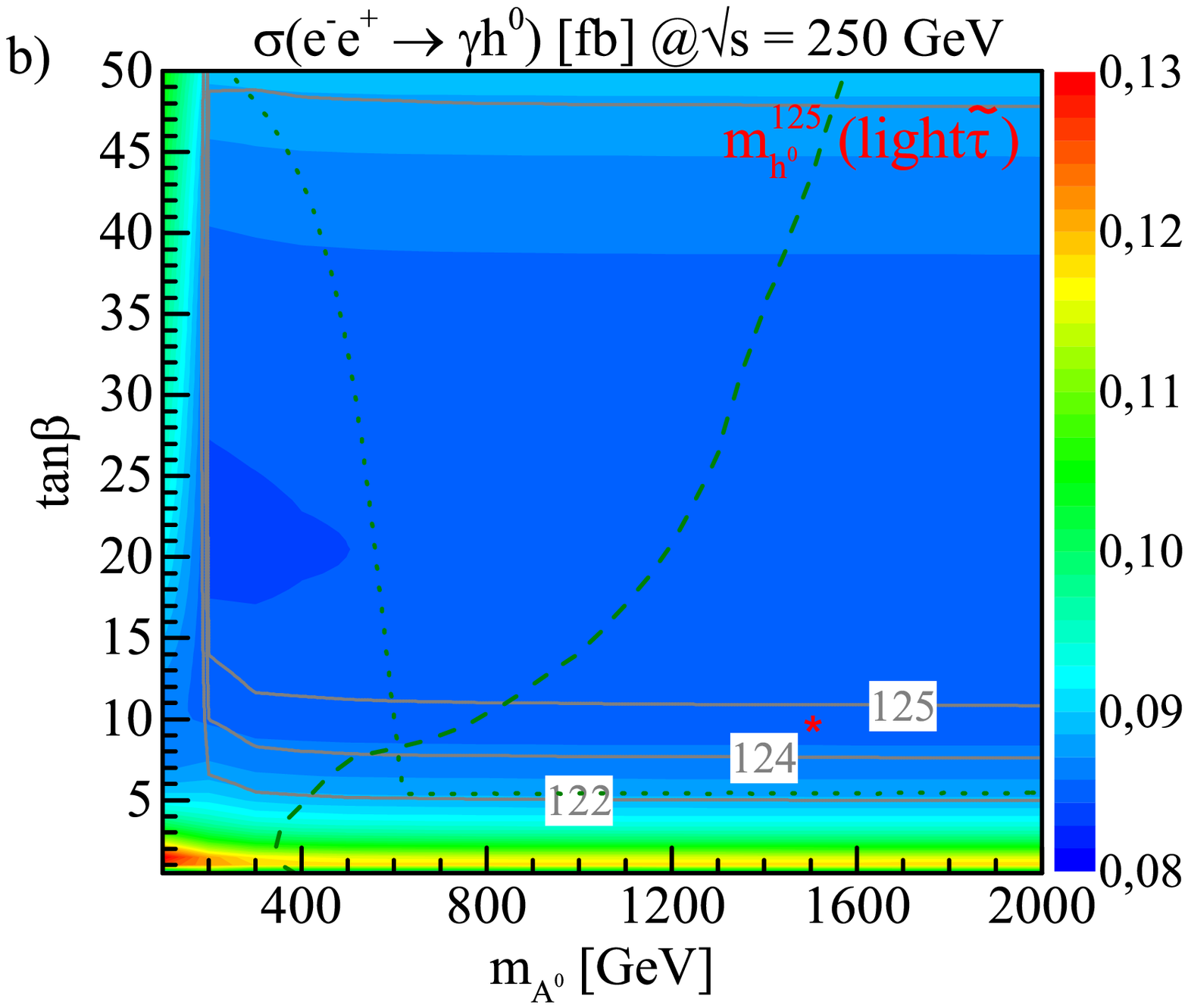}
\includegraphics[scale=0.44]{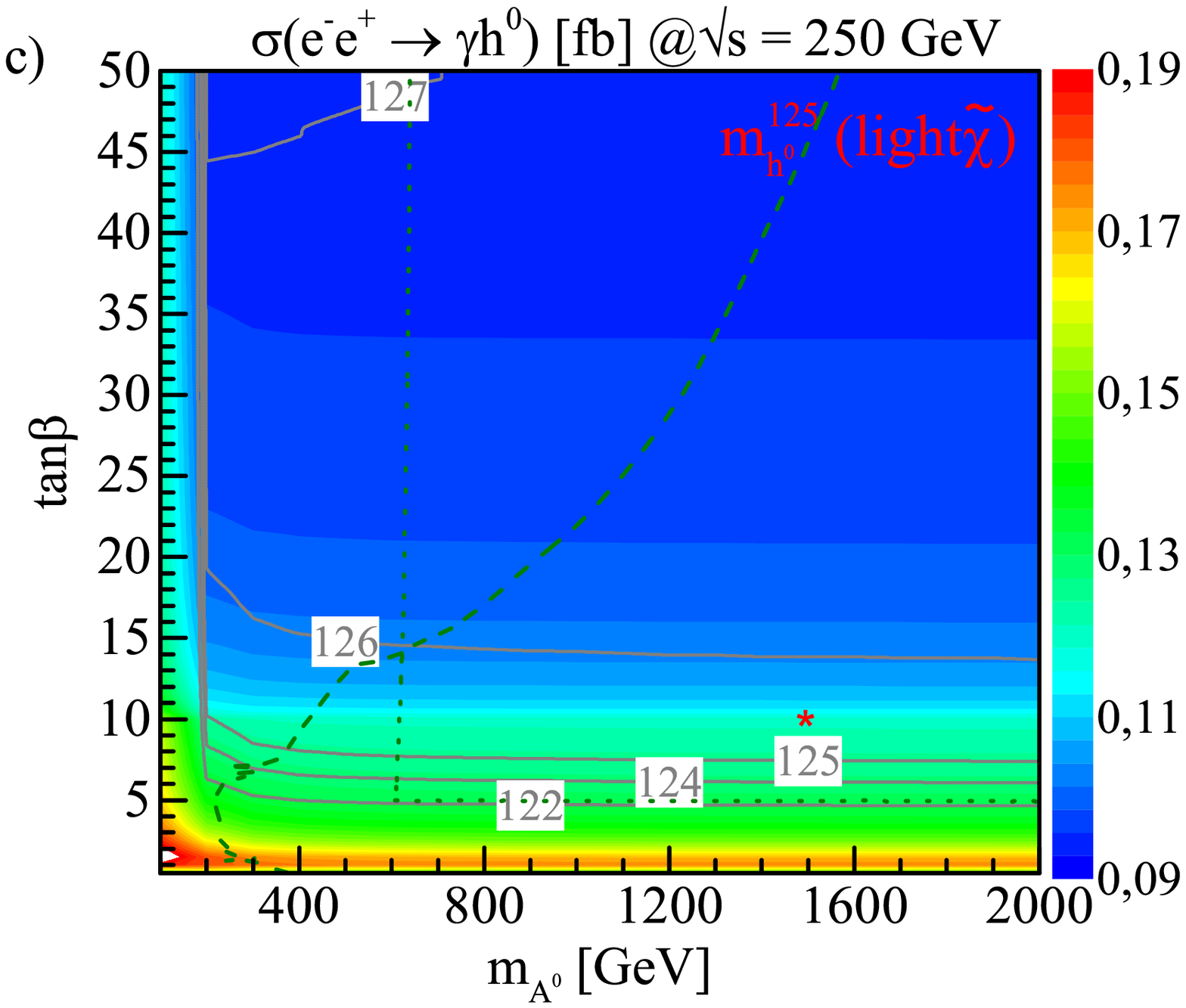}
\includegraphics[scale=0.44]{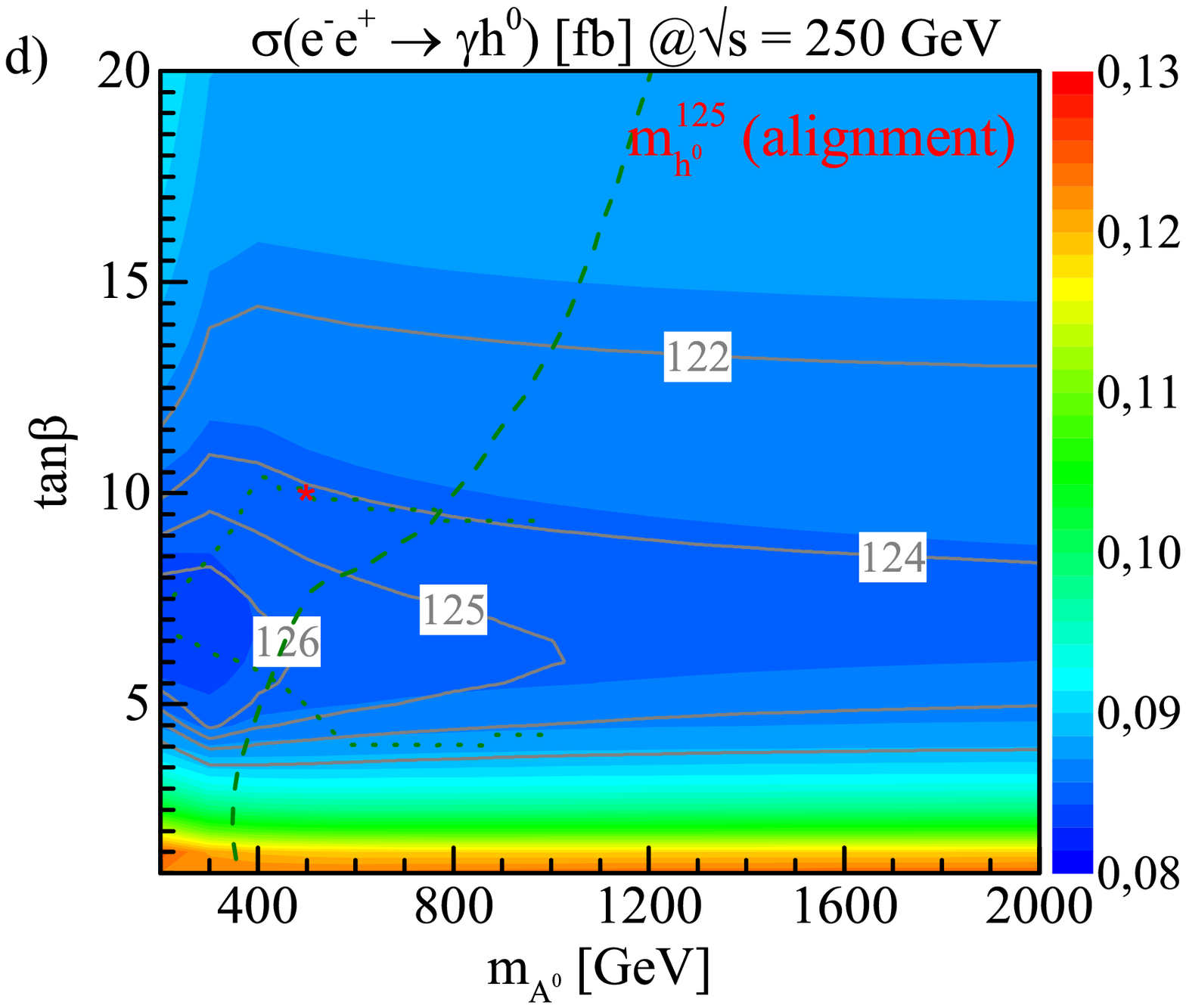}
     \end{center}
\vspace{ -5mm}
\caption{(color online). Total cross section of process $e^- e^+ \to h^0 \gamma$ as a two-dimensional function of $m_A$ and $\tan \beta$ at $\sqrt{s}=250$ GeV for a) the $m_h^{125}$ scenario, b) the $m_h^{125}(light~\widetilde\tau)$ scenario, c) the $m_h^{125}(light~\widetilde\chi)$ scenario, and d) the $m_h^{125}(alignment)$ scenario. The color heat map corresponds to the total cross section (in fb) in the scan region. The gray solid lines indicate predictions for the mass of the light \textit{CP}-even higgs boson $h^0$. The green dashed lines indicate the boundary of the area excluded by the searches for additional Higgs bosons at the LHC. The green dotted lines show the boundary of region excluded by a mismatch between the properties of the light \textit{CP}-even higgs boson $h^0$ and those of the observed Higgs boson. The red stars denote the corresponding benchmark points from the last two rows in Table~\ref{tab:scenarios}.}
\label{fig:hgammaTBMA}
\end{figure*}
It is well known that the total cross section of $e^- e^+ \rightarrow \gamma h^0$ depends on couplings of the Higgs to other particles and  masses of corresponding particles. All the couplings of the Higgs bosons to gauge bosons, fermions, and Higgs bosons are determined by the parameters $m_A$ and $\tan\beta$. The regions of the parameter space where the enhancement of the cross section is large enough to be detectable at a future collider can be found by the behavior with these parameters. In this context, the total cross section of $e^- e^+ \rightarrow \gamma h^0$ is scanned over the plane of $m_A-\tan\beta$ at $\sqrt{s}=250\gev$ as depicted in Fig.~\ref{fig:hgammaTBMA}. The parameters $m_A$ and $\tan\beta$ are varying in the ranges of $100\gev \leq m_A \leq2\tev$ and $0.5 \leq \tan\beta \leq 50$ for the $m_h^{125}$, $m_h^{125}(light~\widetilde\tau)$, and $m_h^{125}(light~\widetilde\chi)$ scenarios and $200\gev \leq m_A \leq2\tev$ and $1 \leq \tan\beta \leq 20$ for the $m_h^{125}(alignment)$ scenario. Additionally, the predictions for the mass of the light \textit{CP}-even Higgs boson $h^0$, $m_h=122\gev,124\gev,125\gev,126\gev$ and $127\gev$, are presented by the contour lines. The corresponding benchmark points are also marked by the red stars. Furthermore, the boundaries of regions excluded by the searches for additional Higgs bosons at the LHC and by a mismatch between the properties of the light \textit{CP}-even higgs boson $h^0$ and those of the observed Higgs boson are shown on parameter space of $m_A$-$\tan\beta$. These boundaries are taken from Ref.~\cite{mh125} for each scenario. It is clear that the total cross section decreases when both $m_A$ and $\tan\beta$ increase for all cases. In particular, the cross section reaches its maximum values at small values of $\tan\beta$ in the scan region.

In the scenario $m_h^{125}$, the predictions for the mass of the light \textit{CP}-even Higgs boson $m_h$ are always below $126 \gev$ all over the plane of $m_A-\tan\beta$ and at $\tan\beta<6$ remain outside the window $125.09\pm 3\gev$ [as shown in Fig.~\ref{fig:hgammaTBMA}{\color{red}(a)}]. At low $m_A$, the decay and production rates of $h$ have been obtained to be incompatible with the LHC results~\cite{mh125}. The total cross section of the production of $e^- e^+ \to \gamma h^0$ reaches about 0.09 fb for the region of $\tan\beta \geq 6$ and any values of $m_A$.

In the scenario $m_h^{125}(light~\widetilde\tau)$, the $m_h$ predictions are smaller than $126 \gev$ all over the plane of $m_A-\tan\beta$, and the smallest value of $\tan\beta$ allowed by the uncertainty $\Delta m_h^{theory}=\pm3\gev$ is around 5 [as shown in Fig.~\ref{fig:hgammaTBMA}{\color{red}(b)}]. The total cross section of the production of $e^- e^+ \to \gamma h^0$ reaches about 0.09 fb for the region of $\tan\beta \geq 5$ and $m_A \geq 200$. Additionally, at large $\tan\beta$ and low $m_A$, the total cross section reaches its biggest values.

In the $m_h^{125}(light~\widetilde\chi)$ scenario, in spite of the decrement in $X_t$, the predictions of $m_h$ display a mild increment with respect to the $m_h^{125}$ scenario. However, these are $m_h<127 \gev$, except for upper-left corner in the plane $m_A-\tan\beta$. The smallest value of $\tan\beta$ allowed by $\Delta m_h^{theory}=\pm3\gev$ is around 5 [as shown in Fig.~\ref{fig:hgammaTBMA}{\color{red}(c)}]. The total cross section of the production of $e^- e^+ \to \gamma h^0$ is about 0.13 fb for the region of $5\leq \tan\beta \leq 8$ and $m_A \geq 200$, and this value decreases with an increasing value of $\tan\beta$.

In the scenario $m_h^{125}(alignment)$, the predictions $m_h$ which are compatible with the measured Higgs mass are placed in the region of $4<\tan\beta<13$ and any values of $m_A$ [as shown in Fig.~\ref{fig:hgammaTBMA}{\color{red}(d)}]. In the allowed parameter region, the cross section reaches about 0.09 fb.

Overall, for production of the \textit{CP}-even Higgs boson $h^0$ in association with a photon, the total cross section could reach a level of $10^{-1}$ fb, depending on the model parameters. This renders the loop-induced process $e^- e^+ \to h^0 \gamma$ in principle observable at a future electron-positron collider. Particularly, FCC$_{\rm ee}$ produces high luminosity for Higgs, $W$, $Z$ and top-quark searches; has multiple detectors; and could reach energies up to the top-pair threshold and beyond. In comparison with other linear $e^- e^+$ colliders such as the ILC and CLIC, the expected luminosity at FCC$_{\rm ee}$ is a factor of between 3 and 5 orders of magnitude larger than that proposed for a linear collider (as shown in Table~\ref{tab:collider}), at all energies from the $Z$-pole to the top-pair threshold, where precision measurements are to be made, hence where the collected statistics will be a key feature. For the expected high luminosity $\mathscr{L}= 10^4$ fb$^{-1}$, the FCC$_{{\rm ee}}$ can provide around two thousand events at $\sqrt{s}=240\gev$. Therefore, the FCC$_{\rm ee}$ can be expected to have a promising potential to detect the process $e^- e^+ \to \gamma h^0$. However, the basic size of the total cross section of $e^- e^+ \to \gamma h^0$ can be enhanced by a factor of 4, depending on the polarizations of the initial $e^-$ and $e^+$ beams. Therefore, the ILC and CLIC, which will be operated with beam polarization, also have an essential role to detect the process $e^- e^+ \to \gamma h^0$. For  $\sqrt{s}=240-250\gev$ and an integrated luminosity of $500$ fb$^{-1}$, an electron-positron collider can produce around $\mathcal{O}(10^5)$ Higgs bosons per year and allow for measuring the Higgs couplings at percent level~\cite{ILC3, Peskin}, which may be able to unravel the MSSM effects in this production.

\begin{figure*}[hpbt]
    \begin{center}
\includegraphics[scale=0.44]{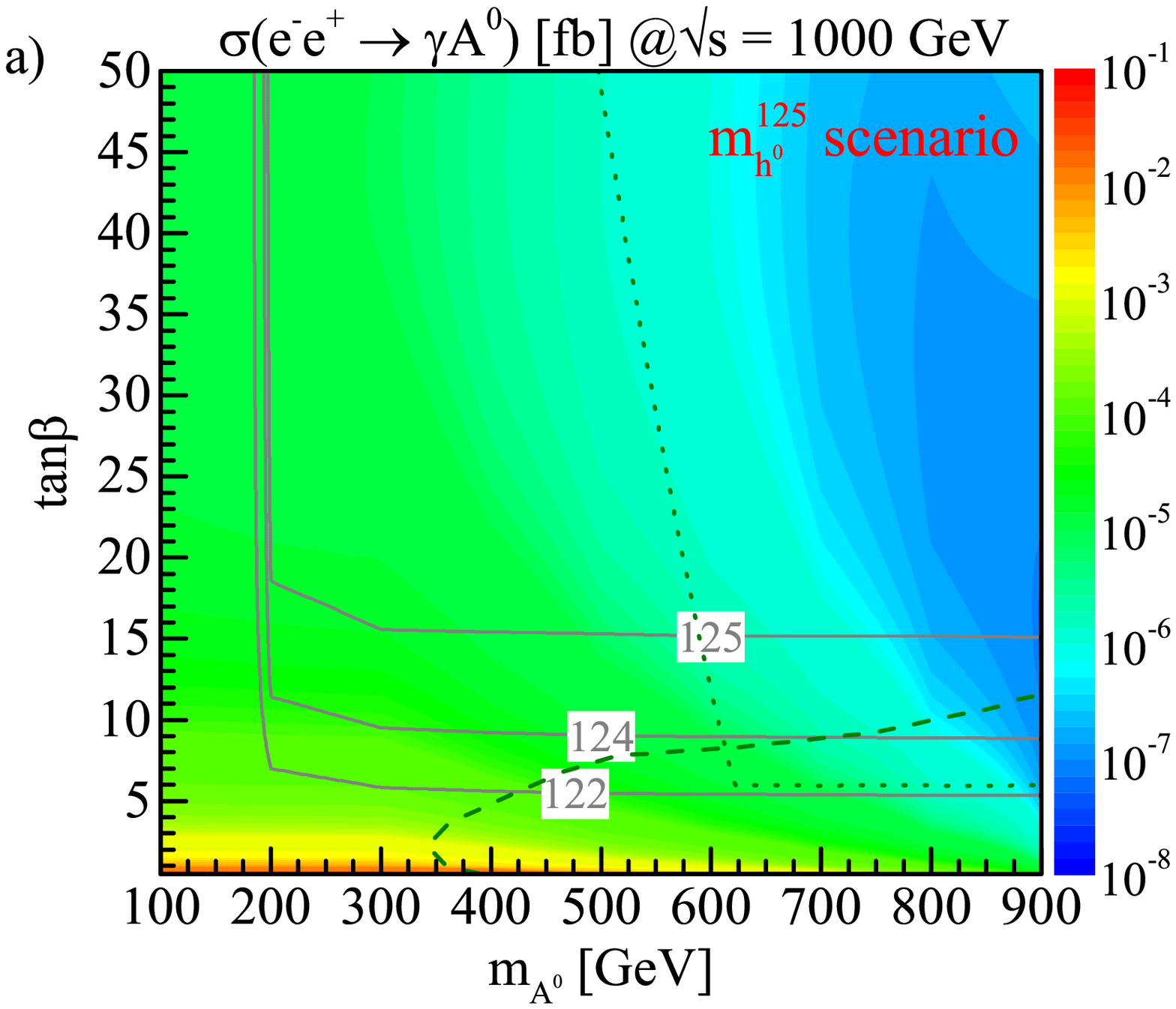}
\includegraphics[scale=0.44]{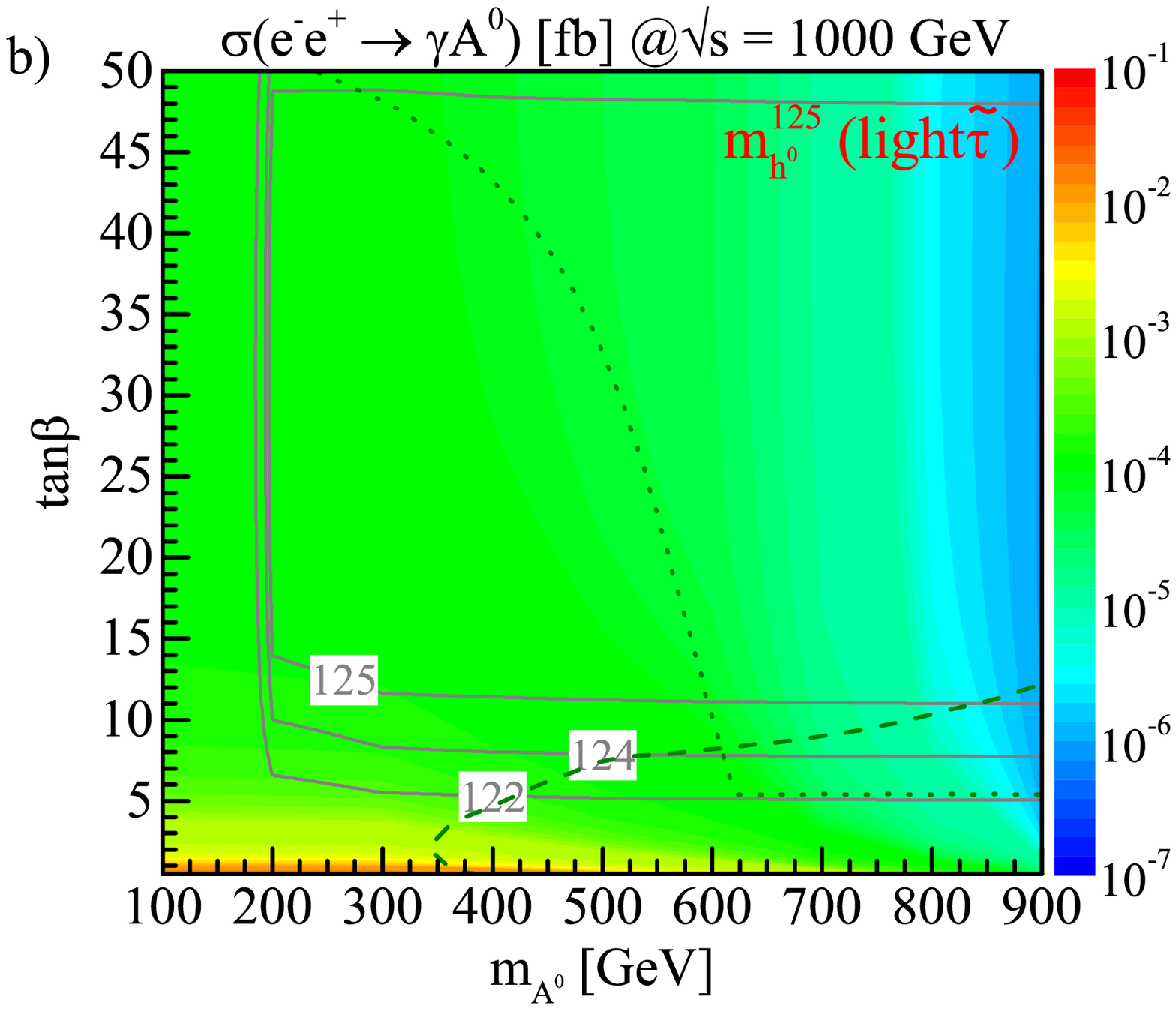}
\includegraphics[scale=0.44]{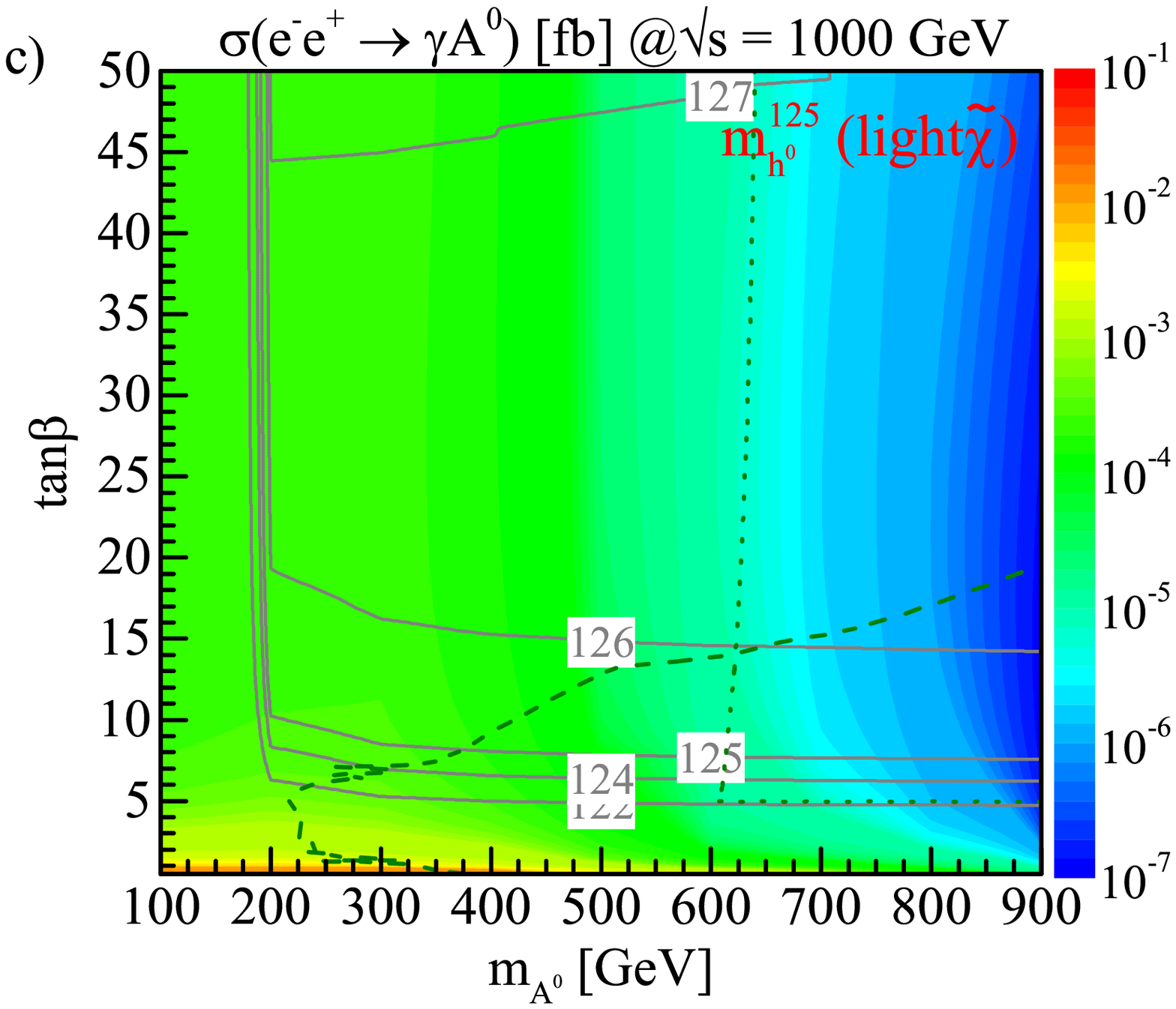}
\includegraphics[scale=0.44]{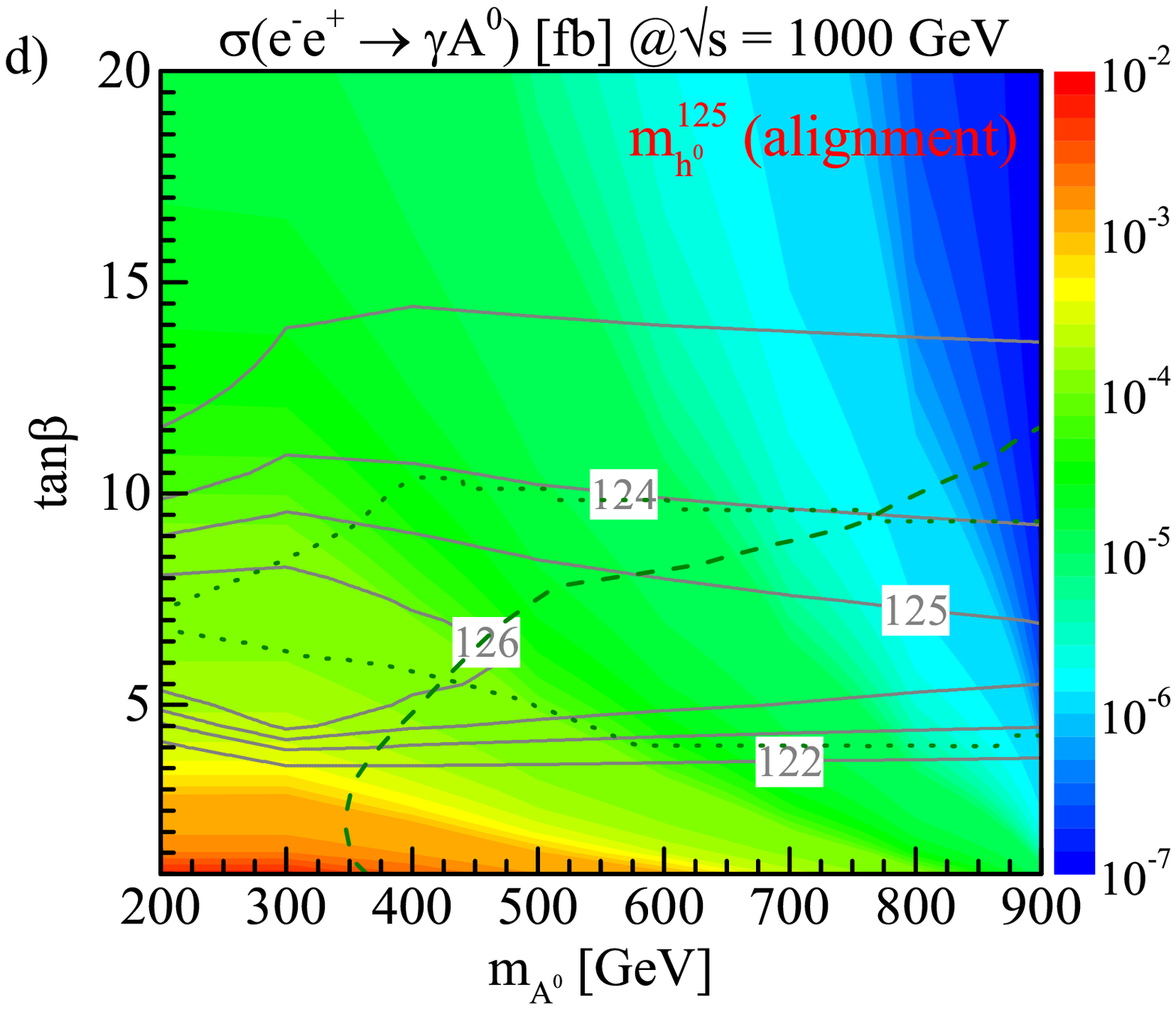}
     \end{center}
\vspace{ -5mm}
\caption{(color online). Total cross section of process $e^- e^+ \to \gamma A^0$ as a two-dimensional function of $m_A$ and $\tan \beta$ at $\sqrt{s}=1$ TeV for each scenario. The color heat map corresponds to the total cross section (in fb) in the scan region. The lines are the same as in Fig.~\ref{fig:hgammaTBMA}.}
\label{fig:AgammaTBMA}
\end{figure*}
\subsection{Process $e^- e^+ \to \gamma A^0$ in the MSSM}
In this study, the single production of the pseudoscalar Higgs boson in association with a photon in electron-positron collisions is also examined in the framework MSSM, considering full one-loop diagrams. The total cross section of $e^- e^+ \rightarrow \gamma A^0$ is scanned over the plane of $m_A-\tan\beta$ at $\sqrt{s}=1\tev$ as depicted in Fig.~\ref{fig:AgammaTBMA}. The parameters $m_A$ and $\tan\beta$ are varying in the range of $100\gev\leq m_A\leq900\gev$ and $0.5 \leq \tan\beta \leq 50$ for the $m_h^{125}$, $m_h^{125}(light~\widetilde\tau)$ and $m_h^{125}(light~\widetilde\chi)$ scenarios and $200\gev \leq m_A \leq900\gev$ and $1 \leq \tan\beta \leq 20$ for the $m_h^{125}(alignment)$ scenario. The total cross section decreases with increments of $m_A$ in all scenarios. The size of the total cross section is at a level of $10^{-2}$ to $10^{-1}$ fb. In particular, the total cross section reaches its largest values at low $m_A$ and $\tan\beta$ into the scan region. There is strong destructive interference between the box-type and triangle-type diagrams, and this reduces the cross section of $e^- e^+ \rightarrow \gamma A^0$. In the scenario $m_h^{125}$, the total cross section of the production of $e^- e^+ \to \gamma A^0$ reaches about 1.22$\times 10^{-4}$ fb for the region of $\tan\beta \geq 6$ and a value of $m_A=400\gev$. For the scenario $m_h^{125}(\widetilde\tau)$, the total cross section reaches about 5.9$\times 10^{-4}$ fb in the region of $\tan\beta \geq 5$ and $m_A = 200\gev$. In the $m_h^{125}(\widetilde\chi)$ scenario, the total cross section is about 7.7$\times 10^{-4}$ fb for $\tan\beta= 5$ and $m_A = 200\gev$, and this value decreases with an increasing value of $\tan\beta$. In the $m_h^{125}(alignment)$ scenario, the total cross section is about 7.9$\times 10^{-5}$ fb for $\tan\beta= 8$ and $m_A = 400\gev$, and this value decreases with an increasing value of $\tan\beta$.

In the allowed parameter space of the considered scenarios,  the size of the total cross section of $e^- e^+ \rightarrow \gamma A^0$ for $\sqrt{s}=1\tev$ is at a level of $10^{-4}$ fb and rather small. This renders the process $e^- e^+ \rightarrow \gamma A^0$ at the border of observability. Consequently, the single production of the neutral Higgs bosons in association with a photon in electron-positron collisions appears to be observable for $\gamma h^0$, but it is very challenging for $\gamma A^0$ at a $e^- e^+$ collider.

\section{Conclusion}\label{sec:conclusion}
In this study, the single production of the neutral Higgs bosons in association with a photon in electron-positron collisions has been analyzed in detail for both the SM and MSSM, focusing on individual contributions from each type of one-loop diagram and polarizations of initial beams. This process has no amplitude at tree level and is hence directly sensitive to one-loop effects and the underlying Higgs dynamics. The four different benchmark scenarios $m_h^{125}$, $m_h^{125}(light~\widetilde\tau)$, $m_h^{125}(light~\widetilde\chi)$, and $m_h^{125}(alignment)$ have been used to illustrate the effect of the new physics in the framework of the MSSM. Note that all of the considered scenarios include a \textit{CP}-even Higgs boson with mass about 125 GeV and couplings consistent with those of the discovered Higgs boson, and a considerable part of their parameter space is allowed by the bounds from the searches for additional Higgs bosons and supersymmetric particles.

Remarkable deviations from the predictions of SM are seen in the $m_h^{125}(light~\widetilde\chi)$ scenario in which the production rate could be significantly enhanced. This scenario induces an enhancement in the production rate up to $58\%$ of that predicted in the SM. In both the SM and MSSM, the cross section is increased up to about four times by the longitudinal polarizations of the initial beams, compared with the unpolarized case. It is clear that the cross section is significantly dependent on the magnitudes of each one-loop amplitude and the relative phases between them in all of the considered scenarios. The $\sigma^{\text{MSSM}}(e^{+} e^{-} \to h\gamma)$ is dominated by the $s$-channel triangle- and the bubble-type contributions at low center of mass energy. Note that the s\_triangle-type amplitude is mainly dominated by the $\tilde{\tau}$ loop because of the $A_{\tau}$ term being large.

The total cross sections of $e^- e^+ \rightarrow \gamma h^0$ and $e^- e^+ \rightarrow \gamma A^0$ were scanned over the plane $m_A-\tan\beta$ for all scenarios. The regions of the parameter space where the enhancement of the cross section is large enough to be observable at a future collider have been presented.

It should be emphasized that precise and model-independent measurements for the single production of the neutral Higgs bosons in association with a photon would be possible at the future $e^{-} e^{+}$ colliders ILC, CLIC, CEPC, and FCC, and therefore the results of this study will be useful in detecting new physics signals based on the MSSM and in providing more precise limits on the corresponding couplings and masses.


\end{document}